\documentclass[aps,pre,reprint,groupedaddress,showpacs]{revtex4-1}

\usepackage{amsfonts} 
\usepackage{amssymb} 
\usepackage{color} 
\usepackage[%
sumlimits,namelimits]{amsmath} 

\usepackage{graphicx}
\usepackage{hyperref}
\usepackage[mathlines]{lineno}

\newcommand{\new}[1]{{\color{red}#1}}
\newcommand{\del}[1]{{\color{red}\sout{#1}}}
\newcommand{\newb}[1]{{\color{blue}#1}}
\newcommand{\delb}[1]{{\color{blue}\sout{#1}}}

\renewcommand{\del}[1]{}
\renewcommand{\new}[1]{{#1}}
\renewcommand{\delb}[1]{}
\renewcommand{\newb}[1]{{#1}}

\begin{document}


\title{Fluctuation Effects in the Pair-Annihilation Process with L\'{e}vy Dynamics}


\author{Ingo Homrighausen}
\altaffiliation{Present address: University of G\"{o}ttingen, Institute for Theoretical Physics, 
Friedrich-Hund-Platz 1, 37077 G\"{o}ttingen, Germany}
\affiliation{Arnold Sommerfeld Center for Theoretical Physics and Center for NanoScience, Department of Physics,\\
Ludwig-Maximilians-Universit\"{a}t M\"{u}nchen, Theresienstra\ss e 37, 80333 M\"{u}nchen, Germany}
\author{Anton A. Winkler}
\affiliation{Arnold Sommerfeld Center for Theoretical Physics and Center for NanoScience, Department of Physics,\\
Ludwig-Maximilians-Universit\"{a}t M\"{u}nchen, Theresienstra\ss e 37, 80333 M\"{u}nchen, Germany}
\author{Erwin Frey}
\affiliation{Arnold Sommerfeld Center for Theoretical Physics and Center for NanoScience, Department of Physics,\\
Ludwig-Maximilians-Universit\"{a}t M\"{u}nchen, Theresienstra\ss e 37, 80333 M\"{u}nchen, Germany}


\date{\today}

\begin{abstract}
We investigate the density decay in the pair-annihilation process $A+A\to\emptyset$ in the case when the particles perform anomalous diffusion on a cubic lattice.
The anomalous diffusion is realized via L\'{e}vy flights, which are characterized by long-range jumps and lead to superdiffusive behavior.
As a consequence, the critical dimension depends continuously on the control parameter of the L\'{e}vy flight distribution.
This instance is used to study the system close to the critical dimension by means of the nonperturbative renormalization group theory.
Close to the critical dimension, the assumption of well-stirred reactants is violated by anticorrelations between the particles, and the law of mass action breaks down.
The breakdown of the law of mass action is known to be caused by long-range fluctuations.
We identify three interrelated consequences of these fluctuations.
First, despite being a nonuniversal quantity and thus depending on the microscopic details, the renormalized reaction rate $\lambda_0$ can be approximated by a universal law close to the critical dimension.
The emergence of universality relies on the fact that long-range fluctuations suppress the influence of the underlying microscopic details.
Second, as criticality is approached, the macroscopic reaction rate decreases such that the law of mass action loses its significance.
And third, additional nonanalytic power law corrections complement the analytic law of mass action term.
An increasing number of those corrections accumulate and give an essential contribution as the critical dimension is approached.
\delb{We demonstrate why these correction terms are more relevant for one-dimensional systems than for two dimensions.}
\newb{We test our findings for two implementations of L\'{e}vy flights that differ in the way they cross over to the normal diffusion in the limit $\sigma\to2$.}
\end{abstract}

\pacs{05.10.Cc, 05.40.-a, 05.40.Fb, 64.60.Ht}


\maketitle


\section{\label{sec:introduction}Introduction}

Reaction diffusion models in arbitrary spatial dimensions $d$ provide a plethora of intensively studied nonequilibrium many body systems~\cite{Lee-1994,*Lee-2006,Lee-1995,Cardy-1998,Vernon-2001}.
One important challenge in studying those systems is to unveil the connection between the microscopic model and the emergent macroscopic physics by integrating out the short-range degrees of freedom.
Intriguingly, there are conditions under which certain macroscopic observables depend only weakly on the microscopic details.
These universal quantities can be observed in low dimensional systems below the critical dimension $d_c$ where long-range fluctuations are significant enough to suppress the microscopic details.
This fact allows one to classify reaction diffusion systems below the critical dimension by universal quantities such as critical exponents~\cite{Odor-2004}.
On the contrary, less effort has been made to investigate reaction diffusion systems above their critical dimension ($d>d_c$).
One reason is that powerful perturbative approaches are conceptually restricted to $d<d_c$ and in general do not allow us to determine the subtle dependence of nonuniversal quantities on the microscopic details in $d>d_c$.
Moreover, since simple mean field approaches provide a qualitatively correct description for $d>d_c$, this regime is considered to be less interesting compared to $d<d_c$.

Despite the applicability of mean field calculations in $d>d_c$, the pair-annihilation process is worth studying above its critical dimension $d_c=2$.
As demonstrated in~\cite{Winkler-2012,*Winkler-2012-b}, the mean field term for three-dimensional pair-annihilation is complemented by an additional nonanalytic correction term.
The correction term originates from long-wavelength fluctuations and is expected to be important close to the critical dimension where the mean field approximation becomes unfaithful.
In this paper we study the pair-annihilation process where the ordinary diffusing Brownian particles are replaced by super diffusing L\'{e}vy flyers.
The motivation to study L\'{e}vy flights is twofold.
First of all, the jump length statistics of a L\'{e}vy flyer leads to superdiffusive behavior, which is argued to be a ubiquitous phenomenon~\cite{Tsallis-1995} and is observed in nature in various contexts, such as turbulence~\cite{Shlesinger-1987,Solomon-1993}, transport phenomena~\cite{Ott-1990,Katori-1997,Zaslavsky-2002,Metzler-2004,Campos-2009}, epidemic spreading~\cite{Janssen-1999,Hinrichsen-2005,Brockmann-2006} or the foraging of animals~\cite{Viswanathan-1999,*Viswanathan-2002,Ramos-2004}.
Besides the practical relevance, superdiffusion has also evolved into a rich field of theoretical research~\cite{Klafter-1987,Fogedby-1994,West-1997,Metzler-2000,Brockmann-2003,Goncharenko-2010}.
Second, from a technical point of view, L\'{e}vy flights allow us to tune the critical dimension and thereby to probe the system close to criticality~\cite{Zumofen-1994}.
This enables us to examine the mechanism of the breakdown of mean field.
We will see that in addition to the correction term known from the diffusive case~\cite{Winkler-2012}, a great number of further corrections become relevant as the critical dimension approaches the spatial dimension from below.
By employing the formalism of nonperturbative renormalization group theory (NPRG), devised by Wetterich et al.~\cite{Wetterich-1991,Wetterich-2002} and elaborated in the context of reaction diffusion models by Delamotte, Canet et al.~\cite{Canet-2004,Canet-2004-B,Canet-2005,Canet-2006,Canet-2011}, the nonuniversal mean field term and the corrections are calculated.

The pair-annihilation process models chemical reactions of two identical particles reacting to an inert product.
The process consists of two competing ingredients.
On the one hand, the erratic motion of the particles leads to the stirring of the reactants.
In the case of L\'{e}vy flights, the erratic dynamics is characterized by the slowly decaying power tail $\propto r^{-d-\sigma}$ according to which the particles perform random jumps of length $r$ (the control parameter $0<\sigma<2$ is referred to as the L\'{e}vy exponent).
On the other hand, the annihilation reaction $A+A\to\emptyset$ tends to build up anticorrelations between the particles and thereby works against the stirring~\cite{Kopelman-1988}.
The critical dimension $d_c$ divides the regime $d>d_c$ where the stirring is dominant such that mean field is valid from the regime $d<d_c$ where long-range anticorrelations lead to fractal-like reaction kinetics and the breakdown of mean field.
As efficient stirring is in favor of the mean field assumption, the critical dimension is lowered for small values of $\sigma$ where the probability for long jumps is high.
This allows us to use the L\'{e}vy exponent $\sigma$ as a control parameter to adjust the critical dimension and probe the system close to criticality.
We will see that $d_c=\sigma$.

For asymptotically late times $t$, the density decay in $A+A\to\emptyset$ is known to obey the power law~\cite{Vernon-2003}
\begin{equation*}
\rho (t)\propto t^{-\alpha},\text{ with }\ \alpha=
\begin{cases}
d/\sigma & \text{for }d<d_c=\sigma \\
1 & \text{for }d>d_c=\sigma.
\end{cases}
\end{equation*}
The results for ordinary diffusion are formally obtained by setting $\sigma=2$.
Instead of looking at the density decay in time, a different, though related question, is to determine the steady state density if the additional particle input $\emptyset\to A$ with rate $J$ is included.
This leads to the scaling law
\begin{equation}
J\propto\rho^\delta,\text{ with }\ \delta=
\begin{cases}
1+\sigma/d & \text{for }d<d_c=\sigma \\
2 & \text{for }d>d_c=\sigma
\end{cases}
\label{eq:steady-state}
\end{equation}
for asymptotically small and homogeneous particle input $J$.
The exponents $\delta$ and $\alpha$ are related by $\delta=1+1/\alpha$~\cite{Racz-1985,*Droz-1993}.
The behavior for $d>d_c$ is correctly predicted by mean field which, in the context of reaction kinetics, is called the law of mass action.
It states that the rate of a chemical reaction is proportional to the concentration of its reactants. 

In the present paper, we focus on the description of the steady state with homogeneous particle input.
Our paper is structured as follows.
After defining the precise mathematical model of anomalous pair-annihilation in Sec.~\ref{sec:model}, we give a brief introduction to the method of nonperturbative renormalization group theory with an emphasize on the application for $A+A\to\emptyset$ in Sec.~\ref{sec:NPRG}.
The results of our analysis are presented in Sec.~\ref{sec:results}.
We conclude in Sec.~\ref{sec:conclusion}.

\section{\label{sec:model}The microscopic model}

As we are going to calculate nonuniversal properties of $A+A\to\emptyset$, we start by giving a detailed definition of the underlying microscopic model.
The $A$ particles are chosen to be represented by idealized, structureless point particles.
In order to provide an UV cutoff, the particles are confined to the sites of some cubic lattice $L=a\mathbb{Z}^d$ with lattice spacing $a$ (the generalization to other lattice geometries is straightforward)
\footnote{
It is also possible to consider extended particles moving in continuous space.
In this case the UV cutoff would be given by the particle extension.
}.
The physical state of the collection of $A$ particles inside the lattice is described by a tuple of non-negative integers $\mathbf{n}=(\dots,\,n_\mathbf{x},\,\dots)$, where $n_\mathbf{x}$ is the occupation number for the lattice site $\mathbf{x}\in L$.

The motion of the particles is modeled by random jumps on the lattice.
Mathematically, this stochastic dynamics is defined as a Markov process, where the random jump events occur independently and Poisson-distributed at rate $1/\tau$.
Hence, the mean time between two consecutive jumps is given by the microscopic time scale $\tau$.
The probability for a particle to jump from lattice site $\mathbf{x}$ to $\mathbf{y}$ \newb{is denoted by $p(\mathbf{x}-\mathbf{y})$ (we assume translational invariance).
The precise form of $p(\mathbf{x}-\mathbf{y})$ is given below.
For the time being, we only require a power law}
\begin{equation}
p(\mathbf{x}-\mathbf{y})
\newb{\simeq \mathcal{A} \left|\mathbf{x}-\mathbf{y}\right|^{-d-\sigma}\text{, for }\left|\mathbf{x}-\mathbf{y}\right|\gg1}
\label{eq:probability}
\end{equation}
with L\'{e}vy exponent $0<\sigma<2$ and amplitude $\mathcal{A}$\delb{ and $\mathcal{N}$ being determined by the normalization condition $\sum_{\mathbf{z}\in L}p(\mathbf{z})=1$}.
\delb{Essential for the L\'{e}vy flight statistics is the slowly decaying power tail in Eq.~(\ref{eq:probability}).}
\newb{The slowly decaying power tail in Eq.~(\ref{eq:probability}) is characteristic of a L\'{e}vy flight statistics.}
More precisely, the second moment of the jump length probability distribution is diverging for $\sigma<2$.
This condition invalidates the central limit theorem and results in superdiffusive dynamics.

The starting point for our calculations is the master equation
\begin{eqnarray}
\partial_t\, P(\mathbf{n},t)=\frac{1}{\tau}\sum_{\mathbf{x},\mathbf{y}\in L} p(\mathbf{x}-\mathbf{y})\ \big[ -n_\mathbf{x}\, P(\mathbf{n},t)\nonumber\\
+(n_\mathbf{x}+1)\,P(\{n_\mathbf{x}+1,n_\mathbf{y}-1\} ,t) \ \big]
\label{eq:master}
\end{eqnarray}
for the probability $P(\mathbf{n},t)$ of finding the system in state $\mathbf{n}$ at time $t$.
The second line in Eq.~(\ref{eq:master}) describes the random jump of one of the $n_\mathbf{x}+1$ particles from site $\mathbf{x}$ to the site $\mathbf{y}$ and is the gain term for the probability flow, whereas the first line constitutes the loss term.
Throughout this paper, we chose units of length and time such that the lattice spacing $a$ and the jump rate $1/\tau$ are equal to $1$.

In the well established procedure devised by several authors~\cite{Doi-1976-I,*Doi-1976-II,Zeldovich-1978,Grassberger-1980,Peliti-1985}, the technique of Fock space formulation is used to map the process (\ref{eq:master}) onto the field theory given by the bare action (space time points are denoted by $x=(t,\mathbf{x})\in\mathbb{R}\times L$)
\[
S_{0}[\bar{\psi},\psi]=\int \text{d}t \sum_{\mathbf{x}\in L} \bar{\psi}_x\,(\partial_t-D_A\nabla^\sigma)\,\psi_x.
\]
Essential for the formalism are the two independent fields $\psi,\bar{\psi}:\mathbb{R}\times L\to\mathbb{R}$ resulting from the coherent state path integral prescription.
Recall that the one-point function $\langle\psi(t,\mathbf{x})\rangle$ is equal to the particle density $\rho(t)$ at time $t$, whereas $\langle\bar{\psi}(t,\mathbf{x})\rangle$ has no direct physical interpretation and is required to vanish due to probability conservation~\cite{Tauber-2005}.
The discrete fractional derivative $-D_A\nabla^\sigma$ encodes the L\'{e}vy flights on the lattice structure $L$ with respect to (\ref{eq:probability}) and is defined by
\[
-D_A\nabla^\sigma\,f(\mathbf{x})\equiv \sum_{\mathbf{z}\in L}p(\mathbf{z})\,[f(\mathbf{x})-f(\mathbf{x}+\mathbf{z})]
\]
In contrast to the discrete Laplacian resulting from the nearest neighbor hopping of an ordinary random walker, the fractional derivative is nonlocal due to the long-range jumps of a L\'{e}vy flyer.
Because of the spatial translational invariance, the operator $-D_A\nabla^\sigma$ is diagonal in Fourier space with integral kernel $\epsilon(\mathbf{p})\equiv 1-\sum_{\mathbf{x}\in L}p(\mathbf{x})\,e^{-i\mathbf{p}\cdot\mathbf{x}}$~\cite{Hinrichsen-1999} and $S_0$ is written as
\begin{subequations}
\label{eq:action}
\begin{equation}
S_{0}[\bar{\psi},\psi]=\int \text{d}t\int_{\mathbf{p}}\bar{\psi}(t,-\mathbf{p})\,\big(\partial_t+\epsilon(\mathbf{p})\big)\,\psi(t,\mathbf{p}),
\label{eq:action-free}
\end{equation}
\end{subequations}
where the integration $\int_{\mathbf{p}}\equiv\int \text{d}^dp/(2\pi)^d$ is performed over the first Brillouin zone $[-\pi,\pi]^d$ of the cubic lattice.
Following the terminology of field theory, we will refer to $\epsilon(\mathbf{p})$ as the dispersion relation.
\newb{
For small momenta $|\mathbf{p}|\ll1$, the dispersion is dominated by the nonanalytic term~\cite{Vernon-2003}}
\begin{equation}
\epsilon(\mathbf{p})\simeq D_A\,|\mathbf{p}|^\sigma,\text{ for }|\mathbf{p}|\ll1.
\label{eq:disp}
\end{equation}
\newb{
The anomalous diffusion constant $D_A$ has dimension of $\text{length}^\sigma/\text{time}$, which indicates superdiffusive scaling between space and time.
Because of its nonanalyticity, the term $D_A\,|\mathbf{p}|^\sigma$ is solely generated by the long-range jumps and can be obtained by approximating the Fourier series in the definition of $\epsilon(\mathbf{p})$ by an integral.
Performing the integration yields the relation between the amplitude $\mathcal{A}$ in Eq.~(\ref{eq:probability}) and the anomalous diffusion constant in Eq.~(\ref{eq:disp})~\cite{Hinrichsen-2005}:}
\begin{equation}
D_A=-\mathcal{A}\,\frac{\pi^{d/2}\,\Gamma(-\sigma/2)}{2^\sigma\,\Gamma\big(\,(\sigma+d)/2\,\big)}.
\label{eq:diff-const}
\end{equation}

\newb{
Besides the requirement of the power law decay for large jumps, we have not specified $p(\mathbf{x})$ yet.
In this paper we want to discuss two different choices for $p(\mathbf{x})$.
First, the pure power law}
\begin{subequations}
\label{eq:probability1}
\begin{equation}
\newb{
p_1(\mathbf{x})
=\mathcal{A}^{(1)}
\begin{cases}
\left|\mathbf{x}\right|^{-d-\sigma} & \text{, }\mathbf{x}\neq0\\
\quad 0 &  \text{, }\mathbf{x}=0,
\end{cases}}
\label{eq:probability1a}
\end{equation}
\newb{
where $\mathcal{A}^{(1)}=1/\sum_{\mathbf{x}\neq 0}|\mathbf{x}|^{-d-\sigma}$ is determined by the normalization condition.
According to Eq.~(\ref{eq:diff-const}), this fixes the diffusion constant to be}
\begin{equation}
\newb{
D_A^{(1)}=-\mathcal{A}^{(1)}\,\frac{\pi^{d/2}\,\Gamma(-\sigma/2)}{2^\sigma\,\Gamma\big(\,(\sigma+d)/2\,\big)}.}
\label{eq:diff-const1}
\end{equation}
\end{subequations}
\newb{The second choice is most easily defined in momentum space ($\mathbf{p}\in[-\pi,\pi]^d$) as}
\begin{equation*}
\newb{
\hat{p}_2(\mathbf{p})\equiv\sum_{\mathbf{x}\in L}p_2(\mathbf{x})\,e^{-i\mathbf{p}\cdot\mathbf{x}}
=\exp(-D_A^{(2)}|\textbf{p}|^\sigma)}.
\end{equation*}
\newb{
(For $\mathbf{p}\in\mathbb{R}^d$, the last expression is the characteristic function of a symmetric L\'{e}vy $\sigma$-stable random variable.)
The inverse Fourier transform of $\hat{p}_2(\mathbf{p})$ leads to the probability density $p_2(\mathbf{x})$ which obeys the required power law decay of Eq.~(\ref{eq:probability}),}
\begin{subequations}
\label{eq:probability2}
\begin{equation}
\newb{
p_2(\mathbf{x})
\simeq \mathcal{A}^{(2)}\,|\mathbf{x}|^{-d-\sigma},\text{ for }|\mathbf{x}|\gg1}
\end{equation}
with 
\begin{equation}
\newb{
\mathcal{A}^{(2)}=-D_A^{(2)}\,\frac{2^\sigma\,\Gamma\big(\,(\sigma+d)/2\,\big)}{\pi^{d/2}\,\Gamma(-\sigma/2)}.}
\label{eq:amplitude2}
\end{equation}
\end{subequations}

\newb{
The qualitative difference between $p_1(\mathbf{x})$ and $p_2(\mathbf{x})$ is manifest in the limit $\sigma\to 2$.
In the case of $p_1$, defined by Eq.~(\ref{eq:probability1}), the amplitude $\mathcal{A}^{(1)}$ remains of $\mathcal{O}(1)$ in the limit $\sigma\to2$.
On the contrary, in the case of $p_2$, the corresponding amplitude $\mathcal{A}^{(2)}$ in Eq.~(\ref{eq:amplitude2}) vanishes linearly in the limit $\sigma\to2$.
A heuristic argument to explain this vanishing is that as $\sigma\to2$, the probability density $p_2$ becomes a Gaussian with an exponentially decaying tail instead of an algebraically decaying power law tail.
The dual observation in momentum space is the following.
While the anomalous diffusion constant $D_A^{(1)}$ in Eq.~(\ref{eq:diff-const1}) diverges linearly in the limit $\sigma\to2$, the diffusion constant $D_A^{(2)}$ in Eq.~(\ref{eq:probability2}) is held constant as $\sigma\to2$.
To summarize, the difference between $p_1$ and $p_2$ in the limit $\sigma\to2$ is that the limit is performed while keeping either the amplitude $\mathcal{A}^{(1)}$ or the diffusion constant $D_A^{(2)}$ fix, respectively.
This difference will turn out to be crucial for the discussion of the two-dimensional pair-annihilation process in Sec.~\ref{sec:results}.
Mathematically, there is no reason why one should favor one of the two possibilities over the other.
It is the physical context that has to determine which model one should use to implement the L\'{e}vy flights.

Let us remark that, whenever we write $p(\mathbf{x})$ and $\epsilon(\mathbf{p})$ without indices, we refer to any probability density with corresponding dispersion relation that fulfill the conditions (\ref{eq:probability}) and (\ref{eq:disp}), respectively.
All formulae written in terms of $p(\mathbf{x})$ and $\epsilon(\mathbf{p})$ hold universally, independent of the specific form of the functions.
We use the index $i\in\{1,2\}$ to indicate that a result may only be valid for a particular choice given by either Eq.~(\ref{eq:probability1}) or Eq.~(\ref{eq:probability2}).
}

Let us come back to the master Eq.~(\ref{eq:master}).
In order to implement that two particles on the same lattice site undergo the pair-annihilation reaction with rate $\lambda$, appropriate terms are added to the master equation which translate to the interaction terms~\cite{Lee-1994}
\begin{equation}
S_{\lambda}[\bar{\psi},\psi]=\int \text{d}t \sum_{\mathbf{x}\in L} \lambda\,(\bar{\psi}_x^2+2\bar{\psi}_x)\,\psi_x^2.
\tag{\ref{eq:action}b}
\label{eq:action-interaction}
\end{equation}
Furthermore, particle input at lattice site $\mathbf{x}$ with rate $J$ is enabled by the term
\begin{equation}
S_{J}[\bar{\psi},\psi]=-J\int \text{d}t \sum_{\mathbf{x}\in L} \bar{\psi}_x.
\tag{\ref{eq:action}c}
\label{eq:action-input}
\end{equation}

In summary, the action functional $S=S_0+S_{\lambda}+S_{J}$ given by Eqs.~(\ref{eq:action-free}) to (\ref{eq:action-input}) provides a field theoretical description of the superdiffusive pair-annihilation process with additional particle input.
As the mapping onto the field theory is exact, all microscopic details (such as the lattice structure) are covered and the description is equivalent to the master equation.

\section{\label{sec:NPRG}The NPRG formalism}

\subsection{General idea}

We are interested in the mean number of particles $\langle n_\mathbf{x}\rangle$ at lattice site $\mathbf{x}\in L$ after time $t$, i.e. we want to compute the particle density
\begin{equation*}
\rho(t,\mathbf{x})\equiv\langle n_\mathbf{x}\rangle\equiv\sum_{\mathbf{n}}n_\mathbf{x}\,P(\mathbf{n},t)
\end{equation*}
at $\mathbf{x}$.
In the field theory formalism, the particle density translates to the one-point function of the $\psi$ field.
With the generating functional $Z[\bar{j},j]=\text{Tr}_{\bar{\psi},\psi}\exp(-S[\bar{\psi},\psi]+\int j\bar{\psi}+\int \bar{j}\psi)$, this reads
\begin{equation}
\rho(t,\mathbf{x})\equiv\langle \psi(x)\rangle\equiv\frac{1}{Z[0,0]}\frac{\delta\,Z[\bar{j},j]}{\delta \bar{j}(x)}\bigg|_{\bar{j}=0=j}\ .
\label{eq:density}
\end{equation}
Assuming a homogeneous initial particle distribution, the mean particle density remains spatially constant, due to the translational invariance of the lattice $L=\mathbb{Z}^d$ and we write $\rho(t,\mathbf{x})=\rho(t)$.

Analogously to Eq.~(\ref{eq:density}), the probability conservation $\sum_{\mathbf{n}}P(\mathbf{n},t)=1$ imposes the constraint $\langle \bar{\psi}\rangle=0$ on the one-point function of the $\bar{\psi}$ field~\cite{Tauber-2005}.

In the saddle point approximation, the dominant contribution to the particle density (\ref{eq:density}) is identified by demanding the phase of the `Boltzmann factor' in $Z$ to be stationary: $\delta\,S/\delta\psi\big|_{\psi=\rho,\bar{\psi}=0}=0=\delta\,S/\delta\bar{\psi}\big|_{\psi=\rho,\bar{\psi}=0}$.
We will refer to this stationary condition as the mean field equation.
If one evaluates the stationary condition for the action functional $S[\bar{\psi},\psi]$ given by Eq.~(\ref{eq:action}), one obtains the mean field rate equation $\partial_t\,\rho=-2\lambda\,\rho^2+J$ for the density.
Despite neglecting the fluctuations around the saddle point, the mean field consideration is known to give a qualitatively correct description above the critical dimension~\cite{Kopelman-1987}.
In fact, the steady state $J=2\lambda\,\rho^2$ yields the correct law of mass action exponent $\delta=2$ for $d>d_c$ (compare Eq.~(\ref{eq:steady-state})).

To go beyond the mean field calculation and thereby to take correlations among the particles into account, one has to compute the generating functional $Z$ beyond the saddle point approximation.
The drawback of the $Z[\bar{j},j]$ functional is that it is not on the same footing as the microscopic action functional $S[\bar{\psi},\psi]$ \new{in the following sense:}
\del{The discrepancy is manifest in two observations.}
\begin{itemize}
\item The action enters $Z$ in the exponent of the `Boltzmann factor' \new{so that $Z$ depends on $S$ in an exponential manner.}
\item The action is a functional of the fields $\psi$ and $\bar{\psi}$, whereas $Z$ depends on the external fields $j$ and $\bar{j}$.
\end{itemize}
Both problems are resolved by defining the generating functional of the connected $n$-point functions $W[\bar{j},j]=\ln Z[\bar{j},j]$ and introducing the \textit{macroscopic (effective) action} $\Gamma[\bar{\psi},\psi]$ to be the functional Legendre transformation of $W$.
As explained below, $\Gamma$ is the generating functional of the (irreducible) vertex functions.
\new{Going from $Z$ to $\Gamma$ preserves all information and $Z$ can be reconstructed from $\Gamma$.
The reason why $\Gamma$ is preferred over $Z$ becomes clear when Eq.~(\ref{eq:density}) and $\langle \bar{\psi}\rangle=0$ are rephrased in terms of the macroscopic action as}
\del{Eq.~(\ref{eq:density}) and $\langle \bar{\psi}\rangle=0$ are rephrased in terms of the macroscopic action as}
\begin{equation}
\frac{\delta\,\Gamma}{\delta\psi}\bigg|_{\psi=\rho,\bar{\psi}=0}=0=\frac{\delta\,\Gamma}{\delta\bar{\psi}}\bigg|_{\psi=\rho,\bar{\psi}=0}\ .
\label{eq:eom-macroscopic}
\end{equation}
This is the fluctuation corrected analog of the mean field equation.
\new{The similarity between Eq.~(\ref{eq:eom-macroscopic}) and the mean field equation suggests to view the functional $\Gamma[\bar{\psi},\psi]$ as the macroscopic counterpart of the microscopic action $S[\bar{\phi},\phi]$.
This indicates that, compared to the $Z$ functional, $\Gamma$ is a more natural quantity to connect the effective (i.e.~fluctuation corrected) physics to the underlying model defined by $S$.
}

We want to calculate the macroscopic action $\Gamma$ based on the underlying microscopic action $S$ by making use of the NPRG method.
In a nutshell, NPRG is based on the idea to incorporate the fluctuations around the saddle point approximation by integrating out degrees of freedom bit by bit in a coarse graining manner.
To this end, a continuous family of effective actions $\Gamma_k[\bar{\psi},\psi]$ is constructed such that the momentum scale parameter $k\in[0,\infty)$ mediates between mean field theory $\Gamma_{k=\infty}[\bar{\psi},\psi]=S[\bar{\psi},\psi]$ and the macroscopic limit $\Gamma_{k=0}[\bar{\psi},\psi]=\Gamma[\bar{\psi},\psi]$.
The construction is carried out by adding the auxiliary mass term
\begin{equation*}
\Delta S_k[\bar{\psi},\psi]=\int \text{d}t\int_{\mathbf{p}}\bar{\psi}(t,-\mathbf{p})\,R_k(\mathbf{p})\,\psi(t,\mathbf{p})
\end{equation*}
to the action $S$.
The function $R_k$ works as a $k$-dependent infrared cutoff and is used to inhibit the propagation of the long range modes in a controlled way.
To leave the physical limit $k=0$ unaffected by this artificial cutoff, one requires $R_0=0$.
On the contrary, for $k=\infty$, one demands $R_{k=\infty}=\infty$ to prevent the propagation of any modes and thereby to freeze all fluctuations in the mean field limit.
Apart from these two constraints, we specify the precise form of $R_k$ below (see Eq.~(\ref{eq:litim})).
The $k$ dependence of $S+\Delta S_k$ is inherited to the generating functional $Z_k$ and leads to the family of effective actions $\Gamma_k$\footnote{An important subtlety is that $\Gamma_k$ is defined as the Legendre transformation of $\ln Z_k$ minus the additional term $\Delta S_k$. This modification of the ordinary Legendre transformation is necessary to establish $\Gamma_{k=\infty}=S$. For $k=0$ we have $\Delta S_k=0$ and the macroscopic action $\Gamma$ is recovered.}.
The effective action $\Gamma_k$ obeys the exact NPRG flow equation
\begin{equation}
\partial_k\, \Gamma_k=\frac{1}{2}\text{Tr}\left[\partial_k\,\tilde{R}_k\left(\Gamma^{(2)}_k+\tilde{R}_k\right)^{-1}\right],
\label{eq:wetterich}
\end{equation}
which establishes the interpolation between the microscopic model and the emerging macroscopic physics.
Equation~(\ref{eq:wetterich}) is referred to as the Wetterich equation and uses an implicit matrix notation such that $\tilde{R}_k$ and $\Gamma_k^{(2)}$ denote the $2\times 2$ matrices of the second functional derivative of $\Delta S_k$ and $\Gamma_k$, respectively.
The quantities that appear under the trace of the Wetterich equation are understood as operators. For example, $\Gamma_k^{(2)}$ is the operator defined through the kernel $\Gamma^{(2)}(x,y)=\delta^2\,\Gamma/\delta\psi(x)\delta\psi(y)$.
The trace is taken with respect to the matrix indices as well as in the operator sense.

Characteristic for the Wetterich equation is its one-loop structure which emerges if it is translated to a Feynman diagram.
In contrast to perturbative treatments, the full field dependent propagator $\big(\Gamma_k^{(2)}[\bar{\psi},\psi]+\tilde{R}_k\big)^{-1}$ is used and renders Eq.~(\ref{eq:wetterich}) to be exact, despite the one-loop structure~\cite{Wetterich-2002}.
A profound consequence of the topological one-loop structure is the fact that both the free part $S_0$ as well as the particle input $S_{J}$ do not become renormalized along the flow (see the following paragraph).
Consequently, we may write the effective action functional as
\begin{equation}
\Gamma_k[\bar{\psi},\psi]=S_0[\bar{\psi},\psi]+S_{J}[\bar{\psi},\psi]+\Gamma_{\text{int},k}[\bar{\psi},\psi]
\label{eq:ansatz}
\end{equation}
with some generic interaction functional $\Gamma_{\text{int},k}[\bar{\psi},\psi]$ that is fixed by the Wetterich flow equation and the initial condition $\Gamma_{\text{int},k=\infty}=S_\lambda$.

\subsection{One-loop structure}

The one-loop structure of Eq.~(\ref{eq:wetterich}) suggests to use diagrammatic techniques in order to exploit the algebraic structure by topological considerations.
Before we begin with the discussion, we give some basic definitions to set up a common language.
Calculations are most efficiently carried out in momentum space where translational invariance is exploited.
We write $p=(\omega,\mathbf{p})$ for the Fourier space dual to $x=(t,\mathbf{x})\in\mathbb{R}\times L$ and abbreviate $\int\frac{d\omega}{2\pi}\int_{\mathbf{p}}$ by $\int_p$.
The vertex functions $\Gamma^{(n,m)}_k(p_i;q_j)$ are defined as the coefficients of the functional Taylor expansion of the interaction functional $\Gamma_{\text{int},k}$ around vanishing fields~\cite{Ryder-1996,*Coleman-1973}:
\begin{eqnarray*}
\Gamma_{\text{int},k}=\sum_{n,m=0}\frac{1}{n!\,m!}\int_{p_1,\dots p_n}\int_{q_1,\dots q_m}\Gamma_k^{(n,m)}(p_i;q_j)\\
\bar{\psi}(p_1)\dots \bar{\psi}(p_n)\cdot\psi(q_1)\dots\psi(q_m).
\end{eqnarray*}
As usual, the vertex functions are represented pictorially by vertices with one leg for each of its arguments, e.g.
\begin{equation*}
\Gamma_k^{(1,2)}(p_1;q_1,q_2)=\parbox{35mm}{\includegraphics[scale=1.0]{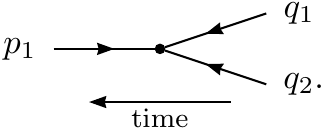}}
\end{equation*}
We discriminate between incoming (right hand side) and outgoing (left hand side) legs which are associated to the $\psi$ and $\bar{\psi}$ fields, respectively.
This notion stems from the fact that the $\psi$ fields are infinitesimally earlier in time than the $\bar{\psi}$ fields, as a result of the path integral construction.

At vanishing fields, the propagator $\big(\Gamma_k^{(2)}[0,0]+\tilde{R}_k\big)^{-1}(p,q)$ follows from (\ref{eq:ansatz}) to be
\begin{equation}
\begin{pmatrix} 0 & \Omega_k(p)^{-1} \\ \Omega_k(-p)^{-1} & 0 \end{pmatrix}(2\pi)^{d+1}\delta(p-q),
\label{eq:propagator}
\end{equation}
where $\Omega_k(p)\equiv -i\omega+\epsilon(\mathbf{p})+R_k(\mathbf{p})$.
Let us emphasize three important consequences of Eq.~(\ref{eq:propagator}).
\begin{itemize}
\item Due to the off-diagonal form, the propagator connects only incoming to outgoing legs.
\item Performing the Fourier transform of (\ref{eq:propagator}) in the temporal direction by using the residue theorem and $\epsilon(\mathbf{p})+R_k(\mathbf{p})>0$ to locate the residues' position relative to the real axis, gives rise to the causality conserving Heaviside step function $\Theta(t_2-t_1)$.
Consequently, the propagator can only connect early $\bar{\psi}$ fields to later $\psi$ fields~\cite{Tauber-2005}.
\item The two points mentioned above, imply that the propagator cannot connect legs of the same vertex~\cite{Canet-2011}.
\end{itemize}
The RG flow of the vertex function $\Gamma_k^{(n,m)}$ is determined by differentiating the Wetterich Eq.~(\ref{eq:wetterich}) with respect to the fields.
In terms of Feynman diagrams, the flow equation is represented by all possible one-loop diagrams with $n$ outgoing and $m$ incoming legs.

A crucial observation is the fact that initially at $k=\infty$ the only nonvanishing vertex functions are $\Gamma_\infty^{(1,2)}$ and $\Gamma_\infty^{(2,2)}$ (see Eq.~(\ref{eq:action-interaction})).
It follows that any vertex which is generated along the renormalization flow cannot have more outgoing than incoming legs~\cite{Peliti-1986}.
As this property holds for $k=\infty$, it suffices to show that the property is preserved by the flow equation, so that it becomes inherited to $k<\infty$:
Since any internal line reduces the number of incoming and outgoing external lines by one, it is impossible to construct one-loop diagrams with more outgoing than incoming external lines and the claim follows.

We conclude that the propagator cannot be dressed by any one-loop diagram.
This instance is special for the pair-annihilation process and is not true in general for other processes such as the branching and annihilating random walk~\cite{Cardy-1998}.
By the same argument, we can rule out the renormalization of the particle input $\Gamma_{J}$ and justify the ansatz (\ref{eq:ansatz}) retrospectively.

The vertex function $\Gamma_k^{(2,2)}$ is of particular significance, as it gives rise to the law of mass action term in the equations of motion (\ref{eq:eom-macroscopic}).
As can be read off from Eq.~(\ref{eq:action-interaction}), the initial vertex function is directly related to the microscopic reaction rate $\lambda$ by $\Gamma_{k=\infty}^{(2,2)}(p_i;q_j)=4\lambda\,(2\pi)^{d+1}\delta(p_i+q_j)$.
The dependence on the momenta is only due to momentum conservation, a result of the point particles' local interaction.
Because momentum conservation is preserved by the propagator (\ref{eq:propagator}), it is inherited to all scales and the most general form for $\Gamma_k^{(2,2)}$ reads
\begin{equation*}
\Gamma_k^{(2,2)}(p_i;q_j)=4\lambda_k(p_i;q_j)\,(2\pi)^{d+1}\delta(p_i+q_j).
\end{equation*}
In order to solve for the running reaction rate $\lambda_k(p_i;q_j)$, we note that the flow equation of $\lambda_k(p_1,-p_1;q_1,-q_1)$ closes, when the total incoming and outgoing momenta vanish separately.
This follows from the simple structure of the Feynman diagram
\begin{equation}
\partial_k\quad \parbox{25mm}{\includegraphics[scale=1.0]{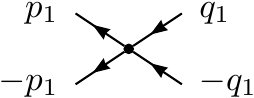}}=\parbox{35mm}{\includegraphics[scale=1.0]{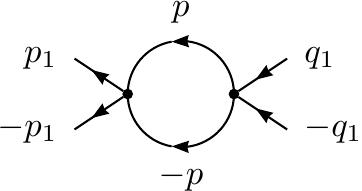}}.
\label{eq:rateflow-diagram}
\end{equation}
Moreover, as $\lambda_k$ is independent of the momenta for $k=\infty$ and the right hand side of Eq.~(\ref{eq:rateflow-diagram}) depends on the external momenta only through the vertex, we conclude inductively, that $\lambda_{k}(p_1,-p_1,q_1,-q_1)=\lambda_k(0,0,0,0)$ is an exact identity on all scales.
Making use of this symmetry and translating the diagrams in (\ref{eq:rateflow-diagram}) into algebraic expressions gives the exact flow equation for $\lambda_k\equiv\lambda_k(0,0,0,0)$:
\begin{equation}
\partial_k\,\lambda_k=\lambda_k^2\int_{\mathbf{p}}\partial_k\,R_k(\mathbf{p})\,\big[\epsilon(\mathbf{p})+R_k(\mathbf{p})\big]^{-2}
\label{eq:rateflow}
\end{equation}
(the loop integration w.r.t.~$\omega_p$ is already performed).
By similar reasoning, we find that $\Gamma_k^{(1,2)}=\Gamma_k^{(2,2)}$, \textit{a priori} only true for $k=\infty$, holds for all $k$.

The flow Eq.~(\ref{eq:rateflow}) can also be obtained by truncating the interaction functional $\Gamma_{\text{int},k}$ to be of the same form as the microscopic interaction $S_\lambda$ (given by Eq.~(\ref{eq:action-interaction})) where $\lambda$ is replaced by the renormalized rate $\lambda_k$.
The reason why this shortcut yields the correct result is that higher vertices, which are created along the renormalization flow, do not contribute to the flow of $\lambda_k$.

\subsection{The cutoff function}

Besides the requirements $R_{k=0}=0$ and $R_{k=\infty}=\infty$ to guarantee $\Gamma_{k=0}=\Gamma$ and $\Gamma_{k=\infty}=S$, respectively, the cutoff function is still unfixed.
A convenient choice for $R_k$ which favors analytic treatment is proposed to be~\cite{Litim-2001,Sengupta-2008} 
\begin{equation}
R_k(\mathbf{p})=\big( \epsilon_k-\epsilon(\mathbf{p})\big)\cdot\Theta\big(\epsilon_k-\epsilon(\mathbf{p})\big).
\label{eq:litim}
\end{equation}
At a given scale $k$, the Heaviside step function separates the physical modes into the low energy (long-ranged) modes with $\epsilon(\mathbf{p})<\epsilon_k$ and the high energy (short-ranged) modes $\epsilon(\mathbf{p})>\epsilon_k$ w.r.t.~the reference energy $\epsilon_k$.
We choose the dependence of the reference energy $\epsilon_k$ on $k$ to approximate the dispersion relation $\epsilon(\mathbf{p})$ for small momenta $|\mathbf{p}|$ (cf. Eq.~(\ref{eq:disp})), i.e., $\epsilon_k=D_A\,k^\sigma$.
Whereas the high energy modes are fully incorporated into $\Gamma_k$ and do not contribute to the flow any longer (due to the term $\partial_k\,\tilde{R}_k$ in Eq.~(\ref{eq:wetterich})), the low energy modes give contributions to the flow.
The prefactor in (\ref{eq:litim}) effects that all low energy modes propagate in the same way:
\begin{equation}
\big(\Gamma_k^{(1,1)}[0,0]+R_k\big)^{-1}(p)=
\begin{cases}
(i\omega+\epsilon_k)^{-1},
& \epsilon(\mathbf{p})<\epsilon_k\\
(i\omega+\epsilon(\mathbf{p}))^{-1}, & \epsilon(\mathbf{p})>\epsilon_k.
\end{cases}
\label{eq:litimpropagator}
\end{equation}
\begin{figure}
\includegraphics{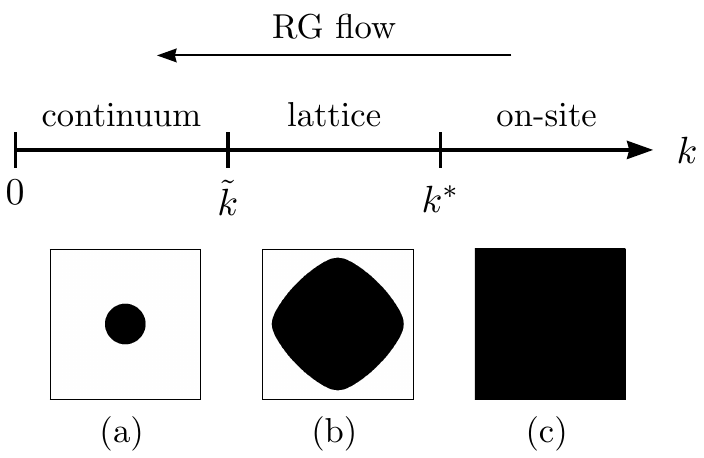}%
\caption{\label{fig:scheme}The RG flow from microscopics ($k=\infty$) to macroscopics ($k=0$) can be divided into three characteristic regimes.
The degrees of freedom being integrated out for $k>k^*$ are very short-ranged and are thus restricted to the lattice sites.
For $\tilde{k}<k<k^*$ the fluctuations propagate through the lattice and thereby perceive the lattice structure.
In the continuum regime $k<\tilde{k}$, the flow is driven by the very long-ranged modes, which cannot resolve the underlying lattice.
The schematic plots below show the typical partition of the two-dimensional Brillouin zone into low (black) and high (white) energy modes for the three characteristic regimes.}
\end{figure}

A crucial consequence of the cutoff (\ref{eq:litim}) is the emergence of two characteristic scales $\tilde{k}$ and $k^*$ dividing the flow into the three parts $k>k^*$, $\tilde{k}<k<k^*$ and $k<\tilde{k}$ (see Fig.~\ref{fig:scheme}), also compare~\cite{Sengupta-2008}.
Let $k^*$ be defined by $\max_{\mathbf{p}}\epsilon(\mathbf{p})=\epsilon_{k^*}$ such that all $\mathbf{p}$ modes in the Brillouin zone are treated as low energy modes for $k>k^*$ (see Fig.~\ref{fig:scheme}~(c)).
As a consequence, the propagator (\ref{eq:litimpropagator}) is independent of the momentum and becomes proportional to $\delta_{\mathbf{x},\mathbf{y}}$ in position space.
The interpretation is that the fluctuations being integrated out for $k>k^*$ correspond to length scales shorter than the lattice spacing and are therefore confined to the lattice sites.
Referring to~\cite{Sengupta-2008}, we call those fluctuations on-site.
For $\tilde{k}<k<k^*$ the high energy modes corresponding to the boundary of the Brillouin zone (see Fig.~\ref{fig:scheme}~(b)), are already integrated out completely and do not contribute to the RG flow.
This regime is dictated by fluctuations which are still short-ranged but do propagate through the lattice and thereby perceive the local lattice structure.
The fact that the low energy modes form a nonisotropic subset of the Brillouin zone (see Fig.~\ref{fig:scheme}) reflects the anisotropy of the lattice at small distances.
In contrast to $k^*$, the boundary $\tilde{k}$ is not defined sharply.
As a rule of thumb, $\tilde{k}$ is the scale below which the low energy modes form an approximately isotropic subset of the Brillouin zone (see Fig.~\ref{fig:scheme}~(a)).
In more technical words, for $k<\tilde{k}$ the dispersion function can be approximated by $\epsilon(\mathbf{p})\approx \epsilon_{|\mathbf{p}|}= D_A\,|\mathbf{p}|^\sigma$ for the low energy modes, which fulfill $\epsilon(\mathbf{p})<\epsilon_k$.
We refer to this regime as the continuum limit of the flow, because the lattice has no impact on the nonanalytic term $D_A\,|\mathbf{p}|^\sigma$ of the dispersion relation (compare the discussion around Eq.~(\ref{eq:diff-const})).
Consequently, for $k<\tilde{k}$, the flow cannot be distinguished from the continuum model where the lattice $\mathbb{Z}^d$ is replaced by $\mathbb{R}^d$.
The position space interpretation is that the flow is driven by long-range fluctuations that propagate through the whole lattice and cannot resolve the lattice structure.
Universal properties, which are independent of the microscopic details, originate from this regime of the flow.

For future reference, we define
\begin{equation*}
\mathcal{V}(k)=\int_\mathbf{p}\Theta\big(\epsilon_k-\epsilon(\mathbf{p})\big)
\end{equation*}
to be the fraction of the low energy modes inside the Brillouin zone.
As $k\lesssim \tilde{k}$, we obtain the continuum scaling $\mathcal{V}(k)\approx K_d\cdot k^d$, where $K_d$ is the volume of the $d$-dimensional unit ball divided by $(2\pi)^d$ (note that $[K_d]=\text{length}^d$).

\section{\label{sec:results}Results}

\subsection{The macroscopic reaction rate}

We use the cutoff function (\ref{eq:litim}) in the flow equation (\ref{eq:rateflow}) of the reaction rate.
The solution of this differential equation with initial condition $\lambda_{k=\infty}=\lambda$ is given by
\begin{equation}
\frac{1}{\lambda_k}=\frac{1}{\lambda}+\int_{\mathbf{p}}\frac{\Theta(\epsilon_k-\epsilon(\mathbf{p}))}{\epsilon_k}+\int_{\mathbf{p}}\frac{\Theta(\epsilon(\mathbf{p})-\epsilon_k)}{\epsilon(\mathbf{p})}.
\label{eq:runningrate}
\end{equation}
In the macroscopic limit $k=0$ this yields
\begin{equation}
\frac{1}{\lambda_0}=\frac{1}{\lambda}+\int_{\mathbf{p}}\frac{1}{\epsilon(\mathbf{p})}
\label{eq:macro-rate}
\end{equation}
and the microscopic rate $\lambda$ becomes connected to its macroscopic counterpart $\lambda_0$.
The inverse rate $1/\lambda_0$, being the typical time scale for the reaction $A+A\to\emptyset$ to take place, is the sum of two contributions.
First, the summand $1/\lambda$ is interpreted as the duration of the reaction between colliding particles based on the microscopic model.
Second, the additional term $\int1/\epsilon>0$ is a measure for the effect of anticorrelations between the particles and slows the process down.
The second term depends on the microscopic details such as the lattice structure through the dispersion function $\epsilon(\mathbf{p})$ and is thus nonuniversal.
\newb{
If necessary, we write $1/\lambda_0^{(i)}=1/\lambda+\int_{\mathbf{p}}1/\epsilon_i(\mathbf{p})$ in order to indicate that certain results may only hold for $\epsilon_i(\mathbf{p})=1-\hat{p}_i(\mathbf{p})$ given by Eqs.~(\ref{eq:probability1}) and (\ref{eq:probability2}) for $i=1$ and $i=2$, respectively.
}

To elaborate on the effect of correlations, we analyze the integral $\int_{\mathbf{p}}1/\epsilon(\mathbf{p})$.
As the domain of integration is finite, possible divergencies can only originate from the region around $\mathbf{p}=0$.
Using $\epsilon(\mathbf{p})=D_A\,|\mathbf{p}|^\sigma$ for asymptotically small momenta, reveals that the integral is IR divergent for $d<\sigma$ which results in $\lambda_0=0$.
Hence, the vanishing of the macroscopic reaction rate $\lambda_0$ is a consequence of long-range fluctuations and entails anomalous reaction kinetics beyond the law of mass action for $\sigma>d$.
Accordingly, we identify $\sigma$ with the critical dimension $d_c$.

\begin{figure}
\includegraphics[width=0.4\textwidth]{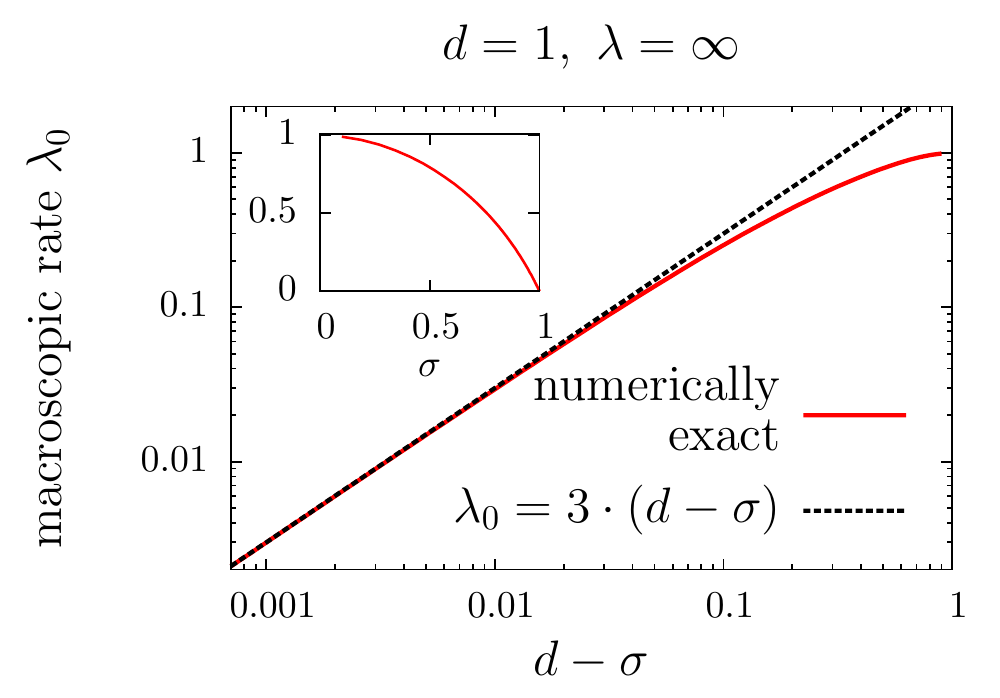}
\caption{\label{fig:powerlaw}
The macroscopic reaction rate (solid red) is plotted against $(d-\sigma)$ for the microscopic rate $\lambda=\infty$ in one spatial dimension.
The dotted black line represents the power law Eq.~(\ref{eq:powerlaw}) which is approached in the limit $(d-\sigma)\to 0$.
The numerical integration of $\int1/\epsilon\newb{_1}$ is carried out by expanding the dispersion function in $\mathbf{p}$ (see Eq.~(\ref{eq:expansion}) of Appendix \ref{sec:expansion}).
The lack of an analogous expansion in two dimensions makes the integration numerically unfaithful for $d=2$.}
\end{figure}

\del{For the sake of concreteness we chose $\lambda=\infty$, such that the macroscopic rate takes the form $1/\lambda_0=\int 1/\epsilon$.}
\new{For the sake of concreteness we assume that particles annihilate instantaneously on contact.
This corresponds to the choice $\lambda=\infty$ and the macroscopic rate takes the form $1/\lambda_0=\int 1/\epsilon$.
As we will explain below, this assumption does not affect our conclusions for the physics close to the critical dimension.}
Figure~\ref{fig:powerlaw} shows the dependence of $\lambda_0$ on the critical dimension $d_c=\sigma<d$ in $d=1$ spatial dimension.
Small values of $\sigma$ correspond to jump length distributions where long jumps occur more frequently and the particles stir more efficiently.
As the stirring works against the correlations, the integral $\int1/\epsilon$ is monotonically increasing in $\sigma$ and as a consequence, $\lambda_0$ decreases with $\sigma$.
For $\sigma=0$ the probability for a jump of length $|\mathbf{x}|$ decays as $|\mathbf{x}|^{-d}$ (see Eq.~(\ref{eq:probability})).
At the same time, the number of possible target sites at the distance $|\mathbf{x}|$ grows as $|\mathbf{x}|^d$.
Consequently, the jump probability is homogeneous in space such that the lattice structure becomes irrelevant and spatial anticorrelations vanish.
Thus, for $\sigma=0$, the effective reaction rate $\lambda_0$ is only limited by the inverse mean time between two consecutive jumps which is set equal to unity.
This explains why the macroscopic reaction rate takes the maximum $\lambda_0=1$ at $\sigma=0$ (see Fig.~\ref{fig:powerlaw}).

In general, the macroscopic reaction rate $\lambda_0$ gets contributions from fluctuations on various length scales as can be seen by the fact that the integral $\int1/\epsilon$ in (\ref{eq:macro-rate}) is performed over the whole Brillouin zone.
Consequently, as already noted above, $\lambda_0$ is nonuniversal and its dependence on the microscopic details is subtle.
Nevertheless, as the critical dimension approaches the spatial dimension from below, the macroscopic rate can be approximated by the universal power law\delb{ (see Fig.~\ref{fig:powerlaw})}
\begin{equation}
\lambda_0\xrightarrow{\sigma\to d=1}\newb{\pi D_A}\cdot(d-\sigma)^1.
\label{eq:powerlaw}
\end{equation}
\new{This equation holds universaly for arbitrary microscopic reaction rates $\lambda$ and not only for $\lambda=\infty$.}
The loss of the nonuniversal character of $\lambda_0$ for $\sigma\to d$ is a consequence of long-range fluctuations.
Mathematically, the reason is that the integral in Eq.~(\ref{eq:macro-rate}) gets its dominant contribution from the small momentum region $|\mathbf{p}|<\tilde{k}\ll 1$, where the dispersion function can be approximated by the continuum limit $D_A\,|\mathbf{p}|^\sigma$.
These long-range fluctuations suppress the nonuniversal, finite contributions arising from $|\mathbf{p}|>\tilde{k}$ in
\begin{equation*}
\frac{1}{\lambda_0}\simeq \int_{|\mathbf{p}|<\tilde{k}}\frac{1}{D_A\,|\mathbf{p}|^\sigma}+\text{finite}
\simeq \frac{1}{\pi\,D_A}\frac{1}{d-\sigma}+\text{finite}
\end{equation*}
so that the dependence of $\lambda_0$ on the short-range physics becomes negligible as $\sigma\to d$.
\new{For $\lambda<\infty$ the nonuniversal finite term also contains the summand $1/\lambda$ from Eq.~(\ref{eq:macro-rate}).}
\delb{The power law (\ref{eq:powerlaw}) is obtained from the last equation because $D_A\xrightarrow{\sigma\to 1}3/\pi$.}
\newb{
The power law (\ref{eq:powerlaw}) holds for both jump length probability distributions, $p_1$ and $p_2$.
In the former case one replaces $D_A$ by $D_A^{(1)}\xrightarrow{\sigma\to 1}3/\pi$ (see Eq.~(\ref{eq:diff-const1})) and obtains $\lambda_0^{(1)}\xrightarrow{\sigma\to d=1}3\cdot(d-\sigma)^1$ (see Fig.~\ref{fig:powerlaw}).
}

The derivation of (\ref{eq:powerlaw}) does not translate \textit{mutatis mutandis} to the two-dimensional case.
\newb{
Instead, we have to distinguish
\begin{subequations}
\label{eq:logarithm}
\begin{equation}
\lambda_0^{(1)}\xrightarrow{\sigma\to d=2}-\mathcal{A}^{(1)}\pi^2/\ln(d-\sigma)
\label{eq:logarithm1}
\end{equation}
and
\begin{equation}
\lambda_0^{(2)}\xrightarrow{\sigma\to d=2}2\pi\, D_A^{(2)}\cdot(d-\sigma)^1,
\label{eq:powerlaw2}
\end{equation}
\end{subequations}
depending on the implementation of the L\'{e}vy flights.
This is the first instance where we encounter the impact of the difference between the jump length probability distributions $p_1$ and $p_2$ on the pair-annihilation process in the limit $\sigma\to2$.
Before we derive Eqs.~(\ref{eq:logarithm1}) and (\ref{eq:powerlaw2}), we want to give a physical explanation.
As noted above, the decrease of the macroscopic reaction rate $\lambda_0$ is a consequence of the anticorrelations among the particles that slow the process down.
Apparently, the decrease in $\lambda_0^{(1)}$ is only logarithmic in the distance to the critical dimension and thus much slower than the linear decrease of $\lambda_0^{(2)}$.
The reason is that $D_A^{(1)}$ diverges as $\sigma\to2$ and thereby increases the anomalous diffusibility of the particles which counteracts the anticorrelations.

The derivation of (\ref{eq:powerlaw2}) is analogous to that of Eq.~(\ref{eq:powerlaw}).
Let us now explain the appearance of the logarithm in Eq.~(\ref{eq:logarithm1}).
}
\delb{
Instead of following a power law, the decay of the macroscopic reaction rate is only logarithmic according to
%
%
where $\mathcal{N}$ denotes the normalization constant of Eq.~(\ref{eq:probability}).
The explanation for the appearance of the logarithm is as follows.}
Naively, applying the same reasoning that leads to (\ref{eq:powerlaw2}) would give $\lambda_0^{(1)}\to2\pi\,D_A^{(1)}\cdot(d-\sigma)$.
However, in the limit $\sigma\to d=2$, the diffusion constant $D_A^{(1)}$ diverges as $D_A^{(1)}\sim 1/(d-\sigma)$ (see Eq.~(\ref{eq:diff-const1})).
Mathematically, the divergence is a consequence of the fact that the anomalous diffusion term $D_A^{(1)}\,|\mathbf{p}|^\sigma$ interferes with the normal diffusion term $D_N^{(1)}\,\mathbf{p}^2$ also contained in $\epsilon_1(\mathbf{p})$.
Both coefficients, $D_A^{(1)}$ and $D_N^{(1)}$, diverge in the limit $\sigma\to2$ such that $D_A^{(1)}/D_N^{(1)}\to-1$ and the sum $D_A^{(1)}\,|\mathbf{p}|^\sigma+D_N^{(2)}\,\mathbf{p}^2$ remains finite.
Consequently, both terms have to be taken into account simultaneously to approximate the small momentum behavior of the dispersion function.
Equation~(\ref{eq:logarithm1}) is then gained by integrating $1/(D_A^{(1)}|\mathbf{p}|^\sigma+D_N^{(1)}\mathbf{p}^2)$ on a small domain around $\mathbf{p}=0$ and expanding the result in $(d-\sigma)$.

A crucial consequence of the decay of $\lambda_0$ is that the law of mass action term $2\lambda_0\rho^2$ is less significant close to the critical dimension.
Correction terms beyond the law of mass action are needed to maintain a quantitatively valid description of the steady state.
The following paragraph explains how the flow of $\lambda_k$ generates nonanalytic corrections to the law of mass action term.

\subsection{\label{sec:corrections}Corrections to the law of mass action}

So far, we have only considered the flow of $\Gamma_k^{(1,2)}$ and $\Gamma_k^{(2,2)}$.
Although initially absent, other vertex functions $\Gamma_k^{(n,m)}$ with $n\leqslant m$ are created along the renormalization flow.
We restrict the vertex functions to be of the form $\Gamma_k^{(n,m)}(p_i;q_j)=n!\,m!\,g_k^{(n,m)}(2\pi)^{d+1}\delta(p_i+q_j)$ with coupling constants $g_k^{(n,m)}\in\mathbb{R}$.
In general, the coupling constants may depend on the external momenta.
In the local potential approximation (LPA) this dependence is neglected so that the interaction functional becomes $\Gamma_{\text{int},k}=\int \text{d}t\sum_{\mathbf{x}\in L}U_k(\bar{\psi},\psi)$ with the local potential $U_k(\bar{\psi},\psi)=\sum_{n,m}g_k^{(n,m)}\,\bar{\psi}^n\psi^m$.
The macroscopic equations of motion~(\ref{eq:eom-macroscopic}) yield
\begin{equation}
J= U_0^{(1,0)}(0,\rho)
\label{eq:eom-potential}
\end{equation}
($U^{(n,m)}$ being the $n$th and $m$th derivative of $U$ w.r.t.~its first and second argument, respectively) for the steady state in the LPA.
As described in the previous section, the leading term of $U_0^{(1,0)}(0,\rho)$ for small densities $\rho$ is the law of mass action term $2\lambda_0\rho^2$.

We are mainly interested in nonanalytic correction terms to the law of mass action.
It is known that the local potential $U_k$ is analytic for $k>0$ and any nonanalytic term can only be created at $k=0$~\cite{Wetterich-2002}.
This puts us in a comfortable position.
In order to determine nonanalytic terms, it is sufficient to investigate the last part of the RG flow, arbitrarily close to $k=0$.
This leads to universality in the sense that nonanalytic terms can only depend on the microscopic model indirectly through the behavior of the RG flow at $k\to 0$.
In more physical words, the nonanalytic terms are solely sourced by long-range modes $\mathbf{p}\simeq 0$ (as $k\to 0$ the cutoff $\partial_kR_k(\mathbf{p})$ in the flow equation suppresses other modes), which cannot resolve the microscopic details such as the lattice structure.
The upshot is that the LPA, despite being an approximation, can be used to compute nonanalytic terms \textit{exactly}, as long as all relevant features of the flow which give rise to the nonanalytic terms are incorporated exactly.

\begin{figure}
\includegraphics[width=0.4\textwidth]{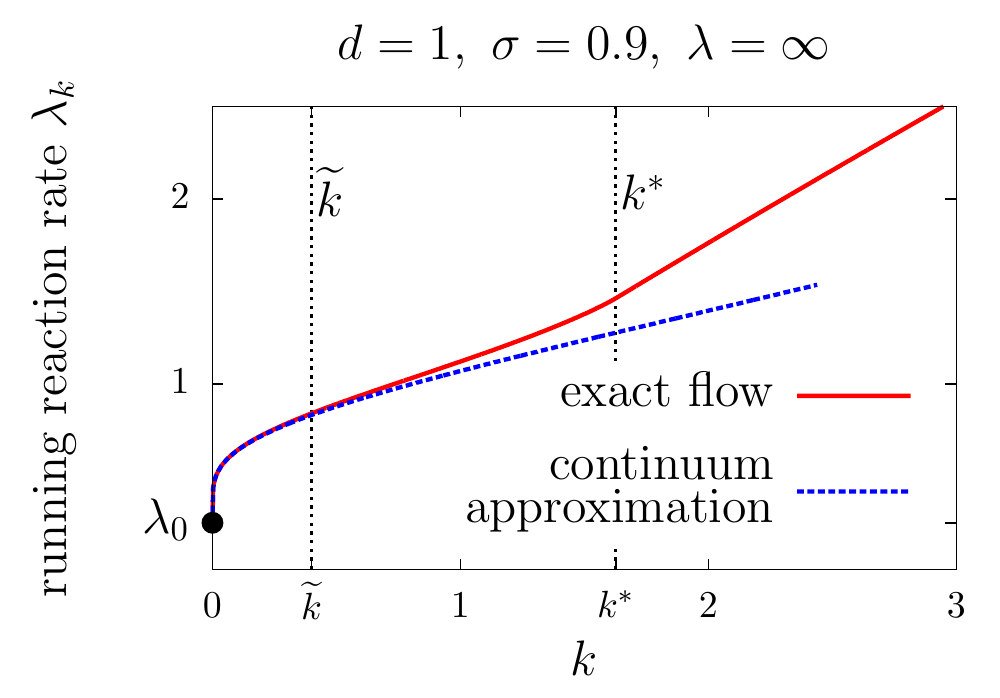}
\caption{\label{fig:rateflow}
The flow of the running reaction rate $\lambda_k$ (solid red) is divided into the three separate regimes $k<\tilde{k}$, $\tilde{k}<k<k^*$ and $k^*<k$ (compare Fig.~\ref{fig:scheme}).
Before the macroscopic limit $\lambda_{k=0}$ is reached (i.e. for $k<\tilde{k}$), the flow is dominated by long-range fluctuations and can be approximated by the continuum limit Eq.~(\ref{eq:rate-continuum}) (dotted blue).
These long-range fluctuations generate nonanalytic corrections to the law of mass action.
As the long-range fluctuations cannot resolve the microscopic details, the nonanalytic terms can only depend on the microscopic model indirectly through the macroscopic reaction rate $\lambda_0$.
Hence, the correction terms are universal functions of the nonuniversal rate $\lambda_0$.}
\end{figure}

We proceed with the computation of nonanalytic correction terms by exploiting the flow equation
\begin{equation}
\partial_k\,U_k^{(1,0)}(0,\rho)=\frac{1}{4}\mathcal{V}(k)\,\partial_k\epsilon_k\,\frac{U_k^{(1,2)}U_k^{(2,0)}(0,\rho)}{\big(U_k^{(1,1)}(0,\rho)+\epsilon_k\big)^2}
\label{eq:LPAflow}
\end{equation}
for the derivative of the local potential.
In terms of Feynman diagrams, the right hand side of Eq.~(\ref{eq:LPAflow}) generates all one-loop diagrams with exactly one outgoing leg.
Almost all of those diagrams are diverging in the limit $k\to 0$.
The divergencies are artificial in the sense that they arise from the fact that one tries to expand the nonanalytic function $U_0^{(1,0)}(0,\psi)$ in the field $\psi$.
Consequently, the diverging Feynman diagrams can be resumed and thereby generate the nonanalytic terms at $k\to 0$.
As can be seen by power counting (more details are given in Appendix \ref{sec:validity}), the most diverging diagrams are the ones that are built out of vertices with two incoming legs only.
This suggests to extract the nonanalytic terms by using the truncation $U_k(\bar{\psi},\psi)=2\lambda_k\bar{\psi}\psi^2+\lambda_k\bar{\psi}^2\psi^2$ in Eq.~(\ref{eq:LPAflow}).
Instead of using the full solution Eq.~(\ref{eq:runningrate}) for $\lambda_k$, the continuum approximation
\begin{equation}
\frac{1}{\lambda_k}=\frac{1}{\lambda_0}-\Sigma_k,\quad\text{where}\quad \Sigma_k=\frac{\sigma}{d-\sigma}\frac{\mathcal{V}(k)}{\epsilon_k}
\label{eq:rate-continuum}
\end{equation}
for small $k$ is sufficient to generate the nonanalytic terms dominated by the long-range physics (see Fig.~\ref{fig:rateflow}).
The term $\Sigma_k$ in Eq.~(\ref{eq:rate-continuum}) measures the deviation of the inverse running coupling $1/\lambda_k$ from its macroscopic limit $1/\lambda_0$.
Applying this ansatz for $U_k$ in Eq.~(\ref{eq:LPAflow}) yields
\begin{equation*}
\begin{split}
&U_0^{(1,0)}(0,\rho)=U_\Lambda^{(1,0)}(0,\rho)\\
&-2\int_0^\Lambda \text{d}k\ \mathcal{V}(k)\partial_k\epsilon_k\left(4+\frac{\epsilon_k}{\lambda_0\rho}-\frac{\epsilon_k\,\Sigma_k}{\rho}\right)^{-2}.
\end{split}
\end{equation*}
The integration may be performed with the help of a computer algebra system after expanding the integrand in powers of $\Sigma_k$.
Due to the universal character, the resulting nonanalytic terms are independent of the arbitrarily chosen integration boundary $\Lambda$.
Eventually, one obtains (exact up to order $\rho^3$)
\begin{subequations}
\label{eq:corrections}
\begin{equation}
J=2\lambda_0\rho^2+\sum_{n\geqslant 1}\mathcal{A}_n\,\rho^{2+n\,\epsilon}
\label{eq:LMA+corr}
\end{equation}
with the amplitudes
\begin{equation}
\begin{split}
\mathcal{A}_n=-&2\lambda_0\, K_d^n\left(\frac{\lambda_0}{D_A}\right)^{n(\epsilon+1)}\\
&4^{n\epsilon}\,\epsilon^{-n+1}\Gamma(-n\epsilon)\frac{\Gamma(1+n\epsilon+n)}{\Gamma(n)}
\end{split}
\label{eq:corr-ampl}
\end{equation}
\end{subequations}
and $\epsilon\equiv \frac{d-\sigma}{\sigma}$ being the relative distance to the critical dimension.
As elaborated in Appendix \ref{sec:validity}, the expression (\ref{eq:corr-ampl}) for the amplitudes $\mathcal{A}_n$ is an exact result close to the critical dimension.

\begin{figure}
\includegraphics[width=0.45\textwidth]{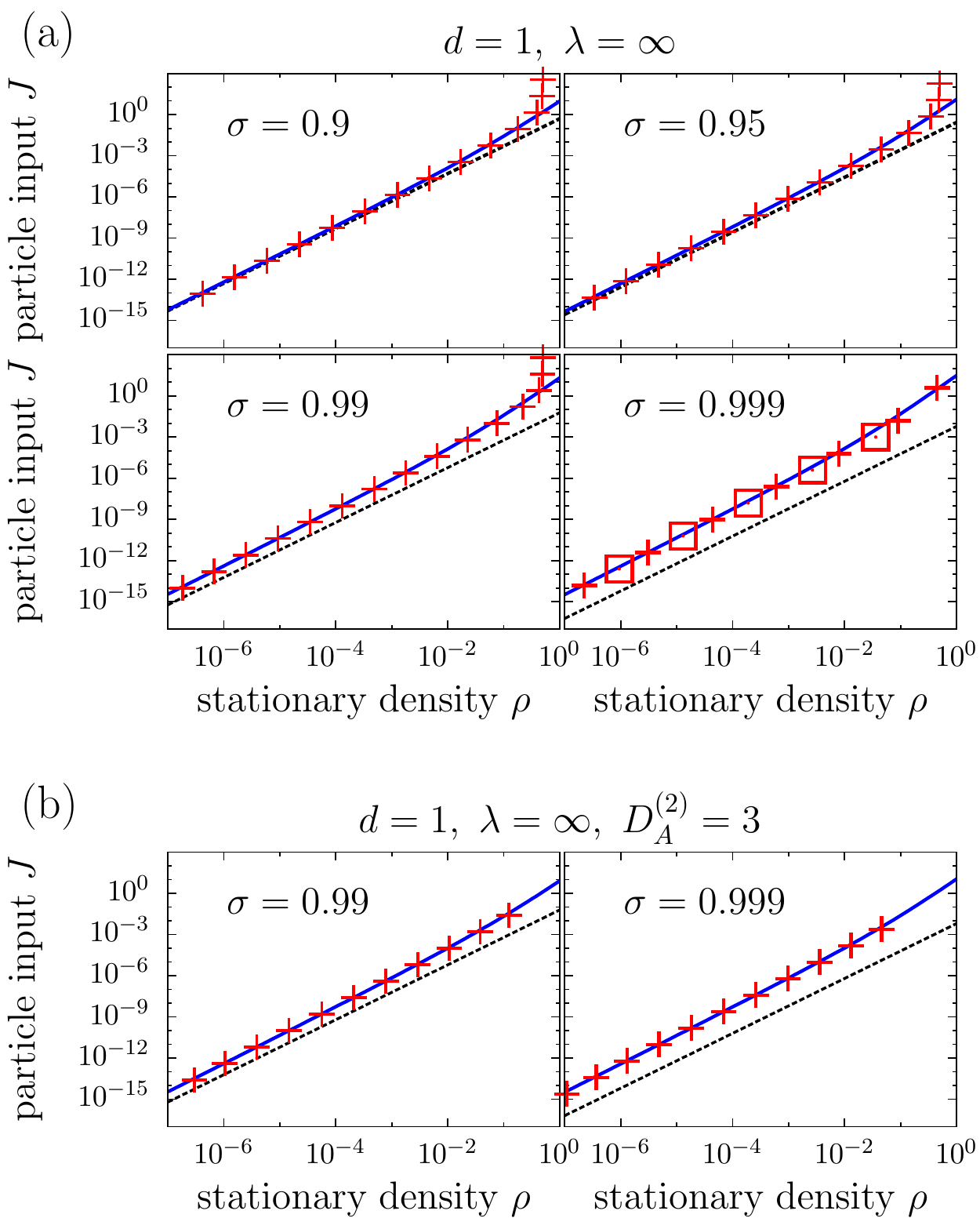}
\caption{\label{fig:simulation}
\newb{(a) }According to the law of mass action, the steady state is determined by $J=2\lambda_0\,\rho^2$ (dotted black line).
As the critical dimension $\sigma$ approaches the spatial dimension $d=1$ from below, the macroscopic reaction rate $\lambda_0$ vanishes according to Eq.~(\ref{eq:powerlaw}) and the law of mass action term loses its significance.
This explains the discrepancy between the simulated data (red crosses) and the theoretical law of mass action (manifest in the last two plots).
Additional nonanalytic correction terms have to be included (blue solid line) in order to compensate the loss of significance of the law of mass action term and to ensure good agreement with the simulations.
\new{The red squares in the lower right plot show the simulation results for finite microscopic reaction rate $\lambda=1$ as opposed to the instantaneous annihilation ($\lambda=\infty$).
Due to the universal behavior of the renormalized rate $\lambda_0$ close to the critical dimension (compare the discussion around Eq.~(\ref{eq:powerlaw})), both data sets collapse on the same curve.}
For this plot, the number of nonanalytic corrections was chosen such that the exponent $2+n\epsilon$ remains below $3$.
\newb{
The L\'{e}vy flights were implemented according to the pure power law $p_1$.
(b) Same plot as in (a), except that the L\'{e}vy flights are now implemented via $p_2$.
The qualitative results are unchanged.}
}
\end{figure}

We give a rough order of magnitude argument to illustrate that the sum of nonanalytic terms in Eq.~(\ref{eq:LMA+corr}) is an essential correction to the law of mass action for $d=1$.
As can be seen in (\ref{eq:LMA+corr}), the exponent $2+n\epsilon$ of the $n$th correction term is close to the analytic law of mass action exponent for small $n\epsilon$.
The corresponding amplitude $\mathcal{A}_n$ is approximated by $\mathcal{A}_n\approx 2\lambda_0\,\left[K_d\lambda_0/(\epsilon\,D_A)\right]^n$.
Using Eq.~(\ref{eq:powerlaw}), we see that the expression in square brackets is approximately one for $d=1$ and $\epsilon\to 0$.
Hence, the amplitudes $\mathcal{A}_n$ are of the same order of magnitude as $2\lambda_0$ and have a similarly small impact as the law of mass action term for $\epsilon\to 0$.
However, as the exponents of the correction terms are close to $2$, many corrections have to be taken into account simultaneously, which leads to an accumulative effect.
If we estimate the number of correction terms which add up coherently by $1/\epsilon$ (such that the exponent $2+n\epsilon$ remains below $3$), we obtain a total contribution of order $1$.
\newb{The crux of the above estimation is the linear decrease of $\lambda_0$ close to the critical dimension, Eq.~(\ref{eq:powerlaw}).
This linear decrease holds for the two L\'{e}vy flight implementations given by $p_1$ and $p_2$.
Hence, as confirmed by computer simulations (see Fig.~\ref{fig:simulation}), the corrections to the law of mass action are important, independent of the implementation of the L\'{e}vy flights.}
\delb{Computer simulations demonstrate the importance of the correction terms (see Fig.~\ref{fig:simulation}).}
\new{A detailed description of the simulation methods can be found in Appendix \ref{sec:simulation}.}

\begin{figure}
\includegraphics[width=0.45\textwidth]{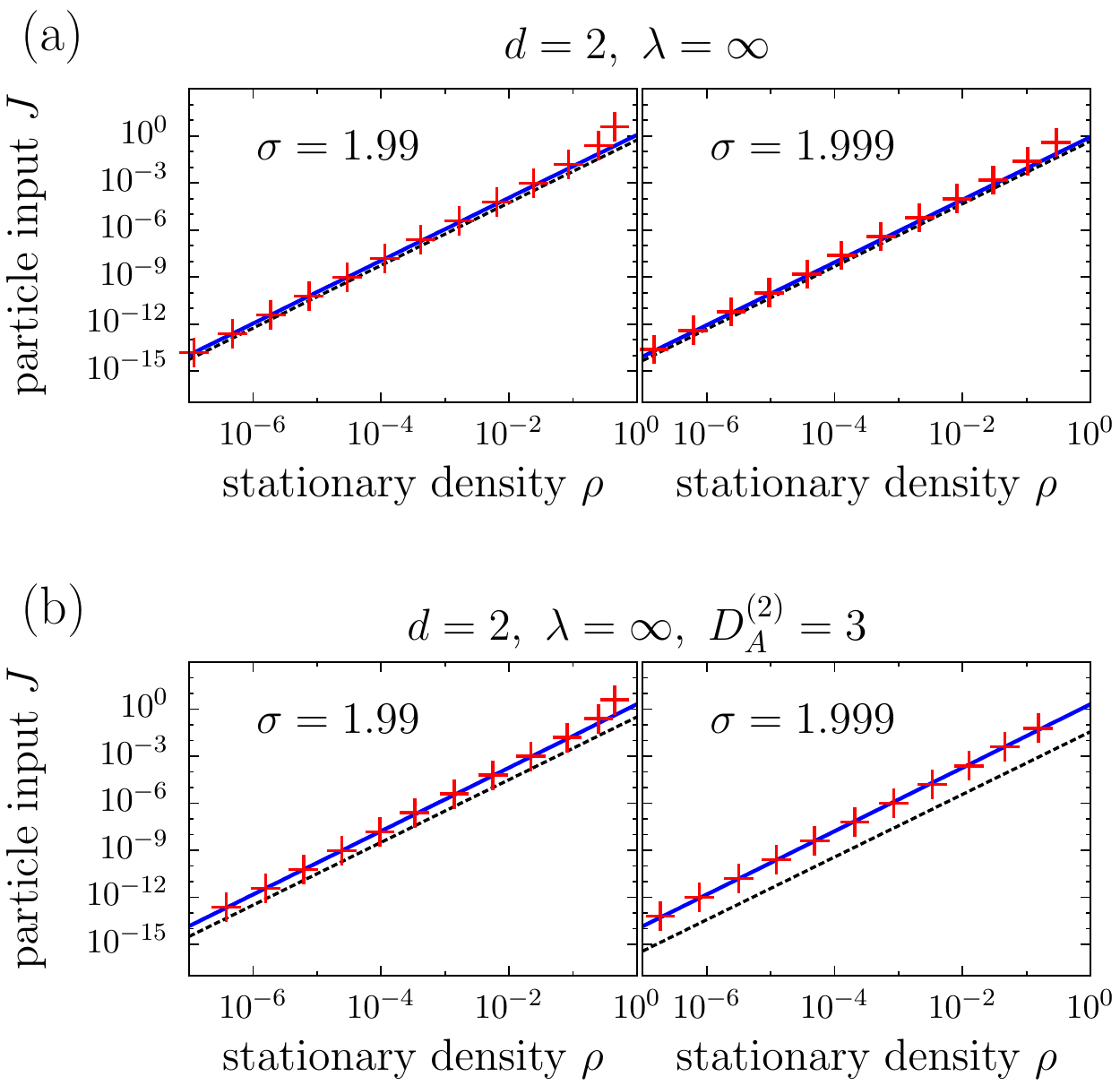}
\caption{\label{fig:simulation2d}
Analogous to Fig.~\ref{fig:simulation} for $d=2$.
\newb{
(a) The L\'{e}vy flights are implemented via the jump length probability distribution $p_1$ and the nonanalytic corrections to the law of mass action term $2\lambda_0\,\rho^2$ are marginal.
(b) If the L\'{e}vy flights are implemented via $p_2$, the correction terms give significant contributions close to the critical dimension.
Mathematically, the difference between (a) and (b) follows from the different behavior of $\lambda_0$ as the critical dimension is approached, see Eqs.~(\ref{eq:logarithm1}) and (\ref{eq:powerlaw2}).}
\delb{In contrast to $d=1$, the nonanalytic corrections to the law of mass action term $2\lambda_0\,\rho^2$ are marginal.
This is related to the fact that the macroscopic coupling $\lambda_0$ decays only logarithmically as the critical dimension $\sigma$ approaches the spatial dimension (see Eq.~(\ref{eq:logarithm})).}
}
\end{figure}

Things are different for $d=2$.
\newb{According to Eqs.~(\ref{eq:logarithm1}) and (\ref{eq:powerlaw2}), the macroscopic reaction rate decreases either logarithmically or linearly, depending on whether the L\'{e}vy flights are implemented via $p_1$ or $p_2$, respectively.
In the latter case, a similar reasoning as in one dimension shows the importance of the correction terms (see Fig.~\ref{fig:simulation2d}(b)).
We make two complementary observations regarding the case when the particle's jump length probability distribution is the pure power law $p_1$.}
First, the divergence of the anomalous diffusion constant $D_A^{(1)}\sim1/(d-\sigma)$ for $\sigma\to 2$ diminishes the amplitudes $\mathcal{A}_n$ given by Eq.~(\ref{eq:corr-ampl}).
This reduces the influence of the correction terms as opposed to the \delb{one dimensional case}\newb{implementation via $p_2$}.
Second, as was pointed out above, the macroscopic reaction rate $\lambda_0^{(1)}$ decays only logarithmically as the critical regime is approached (see Eq.~(\ref{eq:logarithm1}) and the discussion thereafter).
Hence, the law of mass action term $2\lambda_0^{(1)}\,\rho^2$ remains a significant contribution even close to criticality and makes the corrections insignificant.
Our considerations are confirmed by computer simulations shown in Fig.~\ref{fig:simulation2d}.

\section{\label{sec:conclusion}Conclusion}

We have studied the pair-annihilation process with homogeneous particle input of rate $J$ on a $d$-dimensional cubic lattice.
Whenever two particles collide on the same lattice site, they react according to $A+A\to\emptyset$ with rate $\lambda$.
The problem of determining the steady state density $\rho$ for fixed particle input $J$ turns out to be nontrivial.
The main reason for the difficulty is anticorrelations among the particles induced by the annihilation reaction.
In the mean field prescription these anticorrelations are neglected and the steady state is described by $J=2\lambda\,\rho^2$.

The effect of the anticorrelations is to slow down the pair-annihilation reaction and the reaction rate $\lambda$ becomes renormalized.
We have computed the renormalized macroscopic reaction rate $\lambda_0$ exactly and we have found $1/\lambda_0=1/\lambda+\int_{\mathbf{p}}1/\epsilon(\mathbf{p})$.
The crucial point is the integral of the reciprocal of the dispersion function which adds to the inverse microscopic reaction rate and thereby increases the effective reaction time scale.
Consequently, the macroscopic reaction rate is lowered compared to the microscopic rate, which is in agreement with the intuition that anticorrelations slow down the effective reaction.
Because modes of all length scales contribute to the integral, $\lambda_0$ depends on the microscopic details and is thus nonuniversal.

For low dimensional systems below the critical dimension $d_c$, the integral $\int_{\mathbf{p}}1/\epsilon(\mathbf{p})$ is infrared divergent and the macroscopic rate $\lambda_0$ vanishes.
The vanishing of $\lambda_0$ is known as the breakdown of the law of mass action and leads to fractal-like reaction kinetics for $d<d_c$.
In the case where the particles perform L\'{e}vy flights, the critical dimension $d_c$ coincides with the L\'{e}vy exponent $0<\sigma<2$.
\delb{We have used this instance to investigate the system close to criticality.}

\newb{
We have used two different jump length probability distributions $p_1(\mathbf{x}-\mathbf{y})$ and $p_2(\mathbf{x}-\mathbf{y})$ to define the probability that a particle jumps from lattice site $\mathbf{x}$ to site $\mathbf{y}$.
Both distributions obey the slowly decaying power tail $\mathcal{A} |\mathbf{x}-\mathbf{y}|^{-d-\sigma}$ for large jumps of length $|\mathbf{x}-\mathbf{y}|\gg 1$ which is characteristic of L\'{e}vy flights.
Whereas $p_1$ is defined as a pure power law, the definition of $\hat{p}_2$ is reminiscent of the characteristic function of a real-valued L\'{e}vy stable random variable.
This leads to two different implementations of L\'{e}vy flights on the lattice.
The difference is manifest as $\sigma\to2$, i.e.~when the superdiffusive statistics crosses over to normal diffusion.
As explained in this paper, the amplitude $\mathcal{A}$ of the power law tail and the anomalous diffusion constant $D_A$ are Fourier duals of each other, see the discussion around Eq.~(\ref{eq:diff-const}).
It follows that in the limit $\sigma\to2$ only one of the two quantities, either the amplitude or the diffusion constant, can be of order $1$.
In the case of $p_1$, it is the amplitude that is held fixed and the diffusion constant that diverges in the limit $\sigma\to2$.
In the case of $p_2$, the anomalous diffusion constant is held constant and the amplitude vanishes as $\sigma\to2$.
}

The focus of this paper has been on the study of the pair-annihilation process \textit{above} the critical dimension.
\newb{By adjusting the L\'{e}vy exponent and thereby tuning the critical dimension, it is possible to investigate the system close to criticality.}
It turns out, that long-range infrared fluctuations have a significant impact, especially close to the critical dimension.
We have identified three interrelated consequences of these fluctuations which drive the mechanism of the break down of the law of mass action.
First, despite being a nonuniversal quantity and thus depending on the microscopic details, the macroscopic reaction rate $\lambda_0$ can be approximated by a universal law close to the critical dimension.
The emergence of universality relies on the fact that long-range fluctuations suppress the influence of the underlying microscopic details.
Second, as criticality is approached, the macroscopic reaction rate decreases and the law of mass action loses its significance.
\delb{Depending on the spatial dimension, the decrease is either linear or logarithmic for $d=1$ and $d=2$, respectively.}
\newb{
In $d=1$ dimension the decrease is always linear, irrespective of whether $p_1$ or $p_2$ is used.
In $d=2$ dimensions $\lambda_0$ decreases either logarithmically or linearly in the case of $p_1$ and $p_2$, respectively.}
As a consequence, the steady state is no longer accurately described by $J=2\lambda_0\,\rho^2$ close to the critical dimension.
Third, additional nonanalytic power law corrections complement the analytic law of mass action term $2\lambda_0\,\rho^2$.
The corrected steady state equation then reads $J=2\lambda_0\,\rho^2+\sum_n\mathcal{A}_n\,\rho^{2+n\epsilon}$.
As the corrections are generated by the flow of the reaction rate, the amplitudes $\mathcal{A}_n$ are universal functions of the macroscopic rate and have been computed exactly.
When the relative distance $\epsilon=(d-d_c)/d_c$ to the critical dimension is small, the exponents of the nonanalytic terms are close to $2$ and many summands add up coherently.
This compensates the loss of significance of the analytic law of mass action term due to the decay of $\lambda_0$.
\newb{
We have seen that the correction terms are essential in the one-dimensional case.
In two dimensions, one has to distinguish whether the L\'{e}vy flights are implemented via $p_1$ or $p_2$.
Interestingly, the corrections turn out to be less important when the former implementation is used.
}
\delb{Interestingly, the corrections turn out to be less important for $d=2$ dimensions.}
This conforms with the observation that $\lambda_0\newb{^{(1)}}$ decays only logarithmically so that $2\lambda_0\newb{^{(1)}}\,\rho^2$ remains significant which makes the corrections less relevant.
Hence, for practical purposes, the law of mass action remains valid in two dimensions above the critical dimension \newb{as long as the random jumps follow the pure power law distribution $p_1$}.
\delb{In contrast, the correction terms are essential in the one-dimensional case.}
Our theoretical findings are in good agreement with computer simulations.

We have used the method of nonperturbative renormalization group theory (NPRG) to derive our analytic results.
The success of this approach relies on the one-loop structure of the flow equation and the simple initial vertex structure, which is special for the pair-annihilation process.
Accordingly, the propagator does not become renormalized and the simple dependence among the vertex functions enables the exact computation of the running reaction rate $\lambda_k$.
The knowledge of $\lambda_k$ is important for the understanding of the physics driven by the long-range fluctuations close to the critical dimension.
Not only have the long-range fluctuations a crucial impact on the value of $\lambda_0$, but the flow of $\lambda_k$ also generate the corrections to the law of mass action.
The special properties of the pair-annihilation process allowed us to describe the effects of the long-range fluctuations analytically.
But we expect the same qualitative results for more complicated stochastic processes close to criticality.

\begin{acknowledgments}
\newb{
We thank an anonymous referee for setting us on the track to discover the impact of the two different L\'{e}vy flight implementations  for $d=2$ as discussed in the present paper.
}
Financial support of Deutsche Forschungsgemeinschaft through the German Excellence Initiative via the program ``Nanosystems Initiative Munich'' (NIM) and through the SFB TR12 ``Symmetries and Universalities in Mesoscopic Systems'' is gratefully acknowledged.
\end{acknowledgments}

\appendix
\section{\label{sec:expansion}Series expansion of the one-dimensional dispersion function}

We derive the series expansion (\ref{eq:expansion}) of the dispersion function $\epsilon_1(\mathbf{p})$ in $d=1$.
This expansion is essential for the numerical calculation of the integral $\int_{\mathbf{p}} 1/\epsilon_1(\mathbf{p})$ appearing in the macroscopic reaction rate $\lambda_0^{(1)}$ (see Eq.~(\ref{eq:macro-rate})) and has been used to produce Fig.~\ref{fig:powerlaw}.

The definition of the dispersion function in terms of the Fourier transform of $p_1(\mathbf{x})$,
\begin{equation*}
\epsilon_1(\mathbf{p})=1-\sum_{\mathbf{x}\in L} p_1(\mathbf{x})\,e^{-i\,\mathbf{x}\cdot\mathbf{p}}
\end{equation*}
(recall that $p_1(\mathbf{x})$ is the pure power law jump length distribution defined in Eq.~(\ref{eq:probability1a})), has the drawback that the Fourier series converges very slowly for small arguments.
As the small momentum regime gives significant contributions to the macroscopic physics close to the critical dimension, we wish to expand the dispersion function around $\mathbf{p}=0$.
We are able to perform the expansion for the case of a one-dimensional lattice $L=\mathbb{Z}^1$ by doing the summation over the lattice sites:
\begin{equation*}
\begin{split}
\epsilon_1(\mathbf{p})
&=1-2\,\mathcal{N}\ \sum_{c=1}^\infty c^{-1-\sigma}\,\cos(c\,\mathbf{p})\\
&=1-2\,\mathcal{N}\ \sum_{c=1}^\infty c^{-1-\sigma}\,\sum_{n=0}^\infty \frac{(-1)^n}{(2n)!}\,(c\,\mathbf{p})^{2n}\\
&\cong1-2\,\mathcal{N}\ \sum_{n=0}^\infty \frac{(-1)^n}{(2n)!}\,\underbrace{\sum_{c=1}^\infty c^{-1-\sigma+2n}}_{\cong\,\zeta(1+\sigma-2n)}\ \mathbf{p}^{2n}\\[2ex]
&\cong D_A^{(1)}\,|\mathbf{p}|^\sigma-\sum_{n=1}^\infty \frac{(-1)^n}{(2n)!}\,\frac{\zeta(1+\sigma-2n)}{\zeta(1+\sigma)}\ \mathbf{p}^{2n}.
\end{split}
\end{equation*}
The calculation involves three grains of salt, indicated by the modified `equality signs' $\cong$. The naive interchanging (first grain of salt) of the two sums in the third lineleads to the formal expression $\sum_{c=1}^\infty c^{-1-\sigma+2n}$.
Although the sum $\sum_{c=1}^\infty c^{-1-\sigma+2n}$ diverges for $n\geqslant 1$, it is formally substituted by the analytic continuation of the Riemann zeta function $\zeta(1+\sigma-2n)$ (second grain of salt).
A formal power series in $\mathbf{p}$ is obtained.
At the same time, any nonanalytic part of $\epsilon_1(\mathbf{p})$ is lost.
Therefore, the nonanalytic term $D_A^{(1)}\,|\mathbf{p}|^\sigma$ (see Eqs.~(\ref{eq:disp}) and (\ref{eq:diff-const1})) is included by hand in the last line (third grain of salt).

The result
\begin{equation}
\epsilon_1(\mathbf{p})=D_A^{(1)}\,|\mathbf{p}|^\sigma-\sum_{n=1}^\infty \frac{(-1)^n}{(2n)!}\,\frac{\zeta(1+\sigma-2n)}{\zeta(1+\sigma)}\ \mathbf{p}^{2n}
\label{eq:expansion}
\end{equation}
is exact for $\mathbf{p}\in\left[-\pi,\pi\right]$.
Equation~(\ref{eq:expansion}) can be derived mathematically rigorously by exploiting the expansion $\text{Li}_{\sigma+1}(e^p)=\Gamma(-\sigma)\,(-p)^\sigma+\sum_{k=0}\zeta(\sigma+1-k)\,p^k/k!$~\cite{Erdelyi-1953} of the Polylogarithm $\text{Li}_{\sigma+1}(z)\equiv\sum_{k=1}z^k/k^{\sigma+1}$.

We do not know of a generalization of Eq.~(\ref{eq:expansion}) for $d>1$.

\section{\label{sec:validity}On the validity of Eq.~(\ref{eq:corrections})}

The analytic expansion of the local potential reads
\begin{equation*}
U_k=2\lambda_k\bar{\psi}\psi^2+\lambda_k\bar{\psi}^2\psi^2+\sum_{\substack{n\leqslant m\\m\geqslant 3}}g_k^{(n,m)}\bar{\psi}^n\psi^m.
\label{eq:truncation}
\end{equation*}
The condition on the summation indices arises from the fact that no Feynman diagrams with more outgoing than incoming legs are created along the flow.
In view of Eq.~(\ref{eq:eom-potential}), we are interested in the first derivative $U_{k=0}^{(1,0)}(0,\rho)$.
At $k=0$ the potential $U_k$ becomes nonanalytic, and the couplings $g_k^{(n,m)}$ diverge (see Eq.~(\ref{eq:coupling-scaling})).
The proper way to calculate the nonanalytic contributions in $U_0^{(1,0)}(0,\rho)$ is to solve the flow equation~(\ref{eq:LPAflow}) prior to any analytic expansion.
As Eq.~(\ref{eq:LPAflow}) cannot be solved exactly, we have used the truncation
\begin{equation}
U_k=2\lambda_k\bar{\psi}\psi^2+\lambda_k\bar{\psi}^2\psi^2,
\label{eq:truncation}
\end{equation}
for the right hand side of Eq.~(\ref{eq:LPAflow}).
In this proceeding the effect of the higher couplings $g_k^{(n,m)}$ is not neglected completely, since they are generated by $\lambda_k$ in the course of integrating Eq.~(\ref{eq:LPAflow}) w.r.t.~$k$.
It is the back-reaction of $g_k^{(n,m)}$ on the flow of the couplings which is prohibited by the ansatz (\ref{eq:truncation}).
The purpose of this appendix is to show that the back-reaction gives only minor corrections to the amplitudes $\mathcal{A}_n$ in Eq.~(\ref{eq:corrections}) and can be safely neglected.
In particular, the contributions of the back-reaction turn out to be suppressed by the inverse distance to the critical dimension.
The same reasoning can also be applied to justify the momentum independence of $\lambda_k$ in Eq.~(\ref{eq:truncation}).

\begin{figure} 
\includegraphics{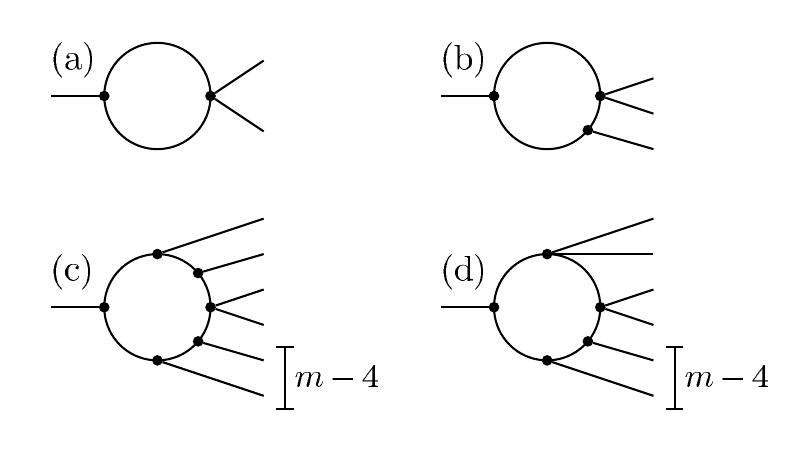}
\caption{\label{fig:diagrams}Some of the one-loop diagrams with one outgoing line are shown.
The most general diagram, which can be built out of vertices with only two ingoing legs is shown in (c).
Diagrams (a), giving rise to the law of mass action term, and (b) are special cases of (c).
Diagram (d) is obtained from (c) by contracting two of the vertices to one vertex with three incoming legs.
As argued in the text, the contribution of (d) is negligible compared to (c).
}
\end{figure}

For the discussion, we use Feynman diagrams as a tool to efficiently organize the power counting in the flow parameter $k$.
This will allow to identify the relevant terms, which generate the nonanalytic corrections.
As already emphasized, the right hand side of Eq.~(\ref{eq:LPAflow}) generates one-loop Feynman diagrams with one outgoing leg and an unrestricted number of incoming legs.
Some of these diagrams are shown in Fig.~\ref{fig:diagrams}.
With the exception of the law of mass action diagram Fig.~\ref{fig:diagrams}(a), all other diagrams lead to divergent terms for $k\to 0$.
To classify the diagrams by their divergence, we apply the following power counting rules.
\begin{itemize}
\item The one-loop structure gives a factor of $\mathcal{V}(k)\to K_d\,k^d$ from the momentum integration and a factor $\partial_k\,\epsilon_k$ from the Wetterich formalism.
\item Each internal line represents the propagator term $1/\big(\epsilon(\mathbf{Q})+R_k(\mathbf{Q})\big)\to 1/\epsilon_k$ as $k\to 0$.
\item Vertices with two incoming legs may be approximated by $\lambda_0$ up to leading order as $k\to 0$.
\end{itemize}
By employing these rules, we find the scaling law $\partial_k\,g_k^{(1,3)}\sim \partial_k\,\epsilon_k\,\mathcal{V}(k)\left(\lambda_0/\epsilon_k\right)^3$ for the diagram of Fig.~\ref{fig:diagrams}(b).
More generally, the diagram of Fig.~\ref{fig:diagrams}(c) with $m$ incoming legs scales as $\partial_k\,\epsilon_k\,\mathcal{V}(k)\left(\lambda_0/\epsilon_k\right)^m$, which results in
\begin{equation}
g_k^{(1,m)}\sim D_A\,K_d\,k^{d+\sigma}\,\left(\frac{\lambda_0}{D_A\,k^\sigma}\right)^m
\label{eq:coupling-scaling}
\end{equation}
for $k\to 0$.
Thus, all couplings beyond the law of mass action diverge in the limit $k\to 0$.
As these divergencies are artifacts of the series expansion in $\rho$, they have to cancel in the sum $J=\lim_{k\to 0}\sum_{m\geqslant 2}g_k^{(1,m)}\rho^m$.
One way to extract finite results is by rewriting the sum in terms of some scaling function $f$ as
\begin{equation}
\sum_{m\geqslant 2} g_k^{(1,m)}\rho^m\sim D_A\,K_d\,k^{d+\sigma}\,f\left(\frac{\lambda_0\,\rho}{D_A\,k^\sigma}\right).
\label{eq:scaling}
\end{equation}
By demanding independence of Eq.~(\ref{eq:scaling}) on the running system size $1/k$ for $k\to 0$, the scaling function is found to behave as the nonanalytic power law $f(x)\sim x^{1+d/\sigma}$ for large arguments $x$.
Apart from numerical factors, this gives the first correction term $\mathcal{A}_1\,\rho^{2+\epsilon}$ of Eq.~(\ref{eq:LMA+corr}).

There are two sources for the higher correction terms $\mathcal{A}_n\,\rho^{2+n\epsilon}$ with $n>1$.
First, the flow of the coupling $\lambda_k$ may be taken into account.
This means that vertices with two incoming legs are not approximated by $\lambda_0$ but rather by
\begin{equation}
\lambda_k=\lambda_0\left(1+\frac{\sigma}{d-\sigma}\frac{\lambda_0\,K_d}{D_A}\,k^{d-\sigma}+\dots\right)
\label{eq:running-lambda}
\end{equation}
(compare Eq.~(\ref{eq:rate-continuum})).
Second, one may include diagrams, which consist of vertices with more than two incoming legs such as the diagram of Fig.~\ref{fig:diagrams}(d).

The truncation given by Eq.~(\ref{eq:truncation}) keeps only diagrams which are built out of vertices with two incoming legs.
To justify this approximation, we argue that diagrams consisting of vertices with more than two incoming legs give only minor corrections to $g_k^{(1,m)}$.
We use the diagram of Fig.~\ref{fig:diagrams}(d) as a concrete example, but the reasoning can also be applied to other diagrams.
Up to leading order in $k$, the vertex with three incoming legs contributes a factor $g_k^{(1,3)}$ given by Eq.~(\ref{eq:coupling-scaling}) for $m=3$.
If we use this scaling to translate the diagram Fig.~\ref{fig:diagrams}(d) into an algebraic expression, we obtain
\begin{equation}
g_{k,\text{corr}}^{(1,m)}\sim g_k^{(1,m)}\,\left(1+\frac{\lambda_0\,K_d}{D_A}\,k^{d-\sigma}+\dots\right)
\label{eq:refined}
\end{equation}
with $g_k^{(1,m)}$ given by Eq.~(\ref{eq:coupling-scaling}).
This is the analog of Eq.~(\ref{eq:running-lambda}).
However, unlike Eq.~(\ref{eq:running-lambda}), the prefactor of the correction term $\sim k^{d-\sigma}$ in Eq.~(\ref{eq:refined}) vanishes as $\lambda_0$ in the limit $\sigma\to d$.
Consequently, we may restrict to diagrams built solely out of vertices with at most two incoming legs and the truncation Eq.~(\ref{eq:truncation}) is justified.

\new{
\section{\label{sec:simulation}Simulation methods}

The purpose of this appendix is to elaborate on the validity of the simulation methods with which Figs.~\ref{fig:simulation} and \ref{fig:simulation2d} were produced.
We have simulated the pair-annihilation process by the well established Gillespie algorithm \cite{Gillespie-1976,*Gillespie-1977}.
Because of the bounded amount of computational resources (especially memory), one can only simulate finite systems.
Hence, we have run the simulations on a finite lattice with periodic boundary conditions.
We denote the total number of lattice sites by $N$.
To rule out finite size effects, we have performed simulations for various system sizes: For $d=1$ we chose $N\in\{10^6,10^7,10^8,10^9\}$ and for $d=2$ we chose $N\in\{300\times300,30\,000\times 30\,000\}$.
The data obtained are insensitive to the system size (see Fig.~\ref{fig:finitesize}).

\begin{figure}
\includegraphics[width=0.45\textwidth]{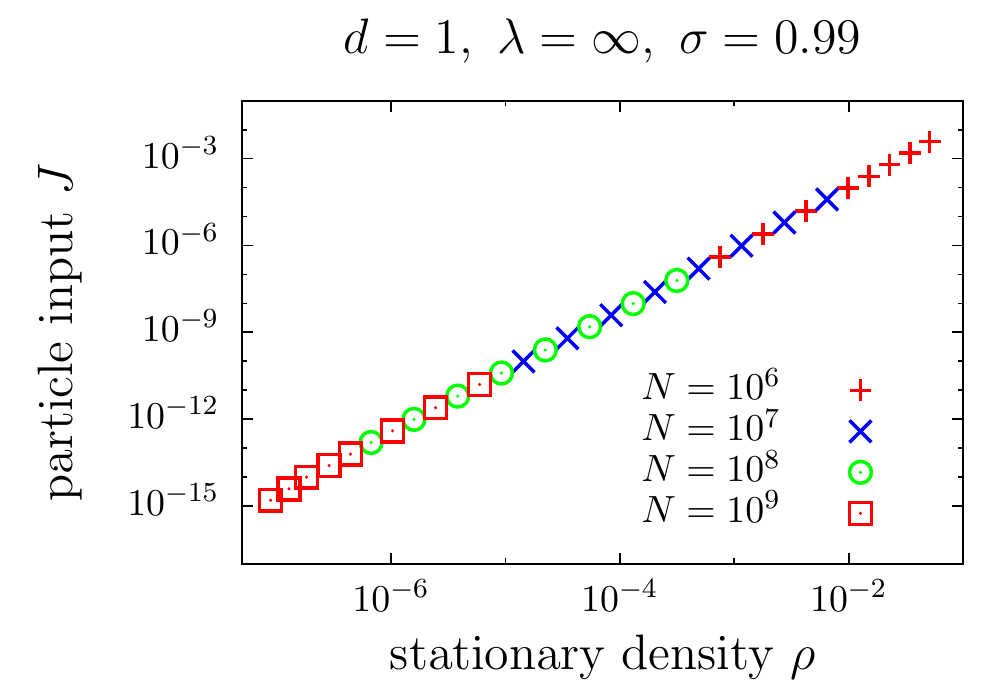}
\caption{\label{fig:finitesize}
The particle input $J$ is plotted against the average density $\rho$ of the stationary state, similar to the plot of Fig.~\ref{fig:simulation}(a).
The different point styles correspond to different choices of the system size $N$.
As all data points clearly lie on the same line, we can exclude finite size effects.}
\end{figure}

Let us comment on the way we have implemented the random jumps according to the probability densities $p_1$ and $p_2$.
For every particle, we have restricted the set of possible target sites to the first $M<N$ neighboring sites only.
This induces a maximal jump length for each particle.
Strictly speaking, because the variance of the jump length distribution is necessarily finite, the particles do not perform superdiffusion but normal diffusion.
However, for reasonable values of $M$ and $\sigma$ the resulting statistics is, for all practical purposes, indistinguishable from the true distribution with diverging variance.
This can be seen by estimating the relative frequency $f$ of jumps larger than the maximal jump length.
For example, in $d=1$ and for $M=10^6$, $\sigma=0.9$ and $p_1(x)=1/(2\zeta(1+\sigma))\, |x|^{-1-\sigma}$, one has $f\approx \int_{\mathbb{R}\backslash [-M/2,M/2]}p_1(x)\,\text{d}x<10^{-5}$.
This means that in less than one out of $10^5$ random jumps a L\'{e}vy flyer might actually `experience' the finiteness of the system.
Table~\ref{tab:table} gives an overview of the values of $M$ used for Figs.~\ref{fig:simulation} and \ref{fig:simulation2d}.
The maximal jump length in two dimensions is significantly smaller than the maximal jump length in $d=1$.
However, as we have investigated two-dimensional systems only for $\sigma>1.9$, large jumps are considerably seldom.
Moreover, note that in the case where the L\'{e}vy flights are implemented via $p_2$, $M$ is a power of $2$.
The reason for this is that we have used the radix-2 Cooley-Tukey algorithm to efficiently transform $\hat{p}_2(\mathbf{p})=\exp(-D_A^{(2)}\,|\mathbf{p}|^\sigma)$ into position space by a fast Fourier transform.
\begin{table}
\caption{\label{tab:table}The values $M$ of possible target sites for a random L\'{e}vy flyer used for Figs.~\ref{fig:simulation} and \ref{fig:simulation2d} is given.
}
\begin{ruledtabular}
\begin{tabular}{c|c|c}
 & $p_1(\mathbf{x})=\mathcal{A}^{(1)}\,|\mathbf{x}|^{-d-\sigma}$ & $\hat{p}_2(\mathbf{p})=\exp(-D_A^{(2)}\,|\mathbf{p}|^\sigma)$\\
\hline
$d=1$ (Fig.~\ref{fig:simulation}) & $M=10^6$ & $M=2^{20}$ \\
$d=2$ (Fig.~\ref{fig:simulation2d}) & $M=4\,000\times4\,000$ & $M=2^{12}\times2^{12}$ \\
\end{tabular}
\end{ruledtabular}
\end{table}
\begin{figure}
\includegraphics[width=0.4\textwidth]{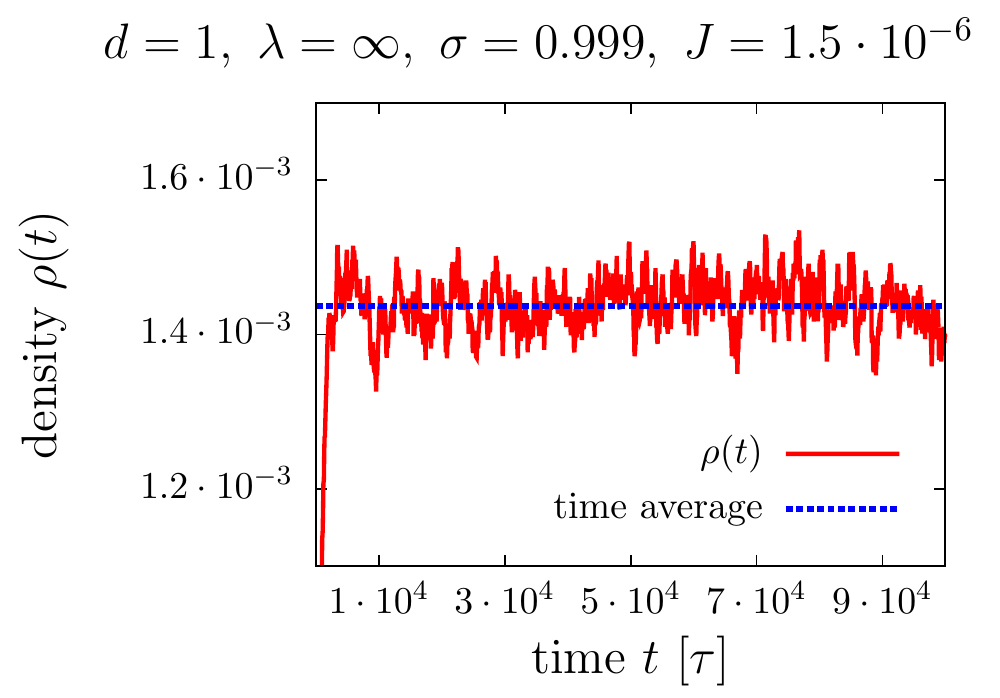}
\caption{\label{fig:gillespie}
The result of a numerical simulation of the pair-annihilation process with particle input $J$ using the Gillespie algorithm on a finite ($N=10^6$) lattice.
Initially, at $t=0$, the lattice is empty.
After a short transient regime the stationary state is reached and the average particle density (dotted blue line) can be measured.}
\end{figure}

We close this appendix by noting that every data point in Figs.~\ref{fig:simulation} and \ref{fig:simulation2d} corresponds to one run of the Gillespie algorithm with fixed particle input $J$ and an initially empty lattice.
After some relaxation time the stationary state is reached and we obtain the mean particle density by averaging the particle density over time.
A typical behavior of the particle density is shown in Fig.~\ref{fig:gillespie}.

}

\providecommand{\noopsort}[1]{}\providecommand{\singleletter}[1]{#1}%

\end{document}